\documentclass[useAMS,usenatbib]{mnras}
\usepackage{graphicx}
\usepackage{amsmath, amssymb}
\usepackage[T1]{fontenc}
\usepackage[usenames]{color}
\usepackage[dvipsnames]{xcolor}
\usepackage{bm}
\usepackage[capitalize]{cleveref}
\usepackage{physics}
\usepackage{bm}
\usepackage{acro}
\usepackage{caption}
\usepackage{subcaption}

\newcommand{\gbar}{g_{\rm bar}}
\newcommand{\gobs}{g_{\rm obs}}
\newcommand{\Vobs}{V_{\rm obs}}
\newcommand{\Jobs}{J_{\rm obs}}
\newcommand{\Vbar}{V_{\rm bar}}

\DeclareRobustCommand{\VAN}[3]{#2}
\let\VANthebibliography\thebibliography
\def\thebibliography{\DeclareRobustCommand{\VAN}[3]{##3}\VANthebibliography}

\newcommand{\Ug}{\Upsilon_{\rm gas}}
\newcommand{\Ud}{\Upsilon_{\rm disk}}
\newcommand{\Ub}{\Upsilon_{\rm bulge}}

\newcommand{\e}{e_{\rm N}}
\newcommand{\gext}{g_{\rm ext}}

\DeclareMathOperator*{\argmax}{arg\,max}

\DeclareAcronym{SPARC}{short = SPARC, long  = Spitzer Photometry \& Accurate Rotation Curves}
\DeclareAcronym{RC}{short = RC, long  = rotation curve }
\DeclareAcronym{RAR}{short = RAR, long  = radial acceleration relation}
\DeclareAcronym{MOND}{short = MOND, long  = Modified Newtonian Dynamics}
\DeclareAcronym{IF}{short = IF, long  = interpolating function}
\DeclareAcronym{EFE}{short = EFE, long  = external field effect}
\DeclareAcronym{ML}{short = ML, long  = machine learning}
\DeclareAcronym{NN}{short = NN, long  = neural network}
\DeclareAcronym{PINN}{short = PINN, long = physics-informed neural network}
\DeclareAcronym{ET}{short = ET, long  = extra-trees}
\DeclareAcronym{RF}{short = RF, long  = random forest}
\DeclareAcronym{XGB}{short = XGB, long  = extreme gradient boosting}
\DeclareAcronym{PC}{short = PC, long  = partial correlation}
\DeclareAcronym{PCA}{short = PCA, long  = principal component analysis}
\DeclareAcronym{LRELU}{short = LReLU, long  = leaky rectified linear unit}
\DeclareAcronym{LCDM}{short = $\Lambda$CDM, long  = $\Lambda$ Cold Dark Matter}
\DeclareAcronym{TFR}{short = TFR, long  = Tully--Fisher relation}
\DeclareAcronym{FP}{short = FP, long  = fundamental plane}
\DeclareAcronym{RARIF}{short = RAR IF, long  = RAR interpolating function}
\DeclareAcronym{SIFEFE}{short = Simple IF $+$ EFE, long  = simple interpolating function $+$ external field effect}
\DeclareAcronym{SHAP}{short = SHAP, long = Shapley additive explanation}
\DeclareAcronym{SDSS}{short = SDSS, long = Sloan Digital Sky Survey}
\DeclareAcronym{RMS}{short = RMS, long = root-mean-square}

\title[Fundamentality of the RAR]{On the fundamentality of the radial acceleration relation for late-type galaxy dynamics}

\author[R. Stiskalek and H. Desmond]{
Richard Stiskalek$^{1}$\thanks{\href{mailto:richard.stiskalek@physics.ox.ac.uk}{richard.stiskalek@physics.ox.ac.uk}}
and Harry Desmond$^{2}$\thanks{\href{mailto:harry.desmond@port.ac.uk}{harry.desmond@port.ac.uk}}
\\
$^{1}$Astrophysics, University of Oxford, Denys Wilkinson Building, Keble Road, Oxford, OX1 3RH, UK\\
$^{2}$Institute of Cosmology \& Gravitation, University of Portsmouth, Dennis Sciama Building, Portsmouth, PO1 3FX, UK
}

\date{Accepted XXX. Received YYY; in original form ZZZ}
\pubyear{2023}

\begin{document}\label{firstpage}
\pagerange{\pageref{firstpage}--\pageref{lastpage}}
\maketitle

\begin{abstract}
Galaxies have been observed to exhibit a level of simplicity unexpected in the complex galaxy formation scenario posited by standard cosmology. This is particularly apparent in their dynamics, where scaling relations display much regularity and little intrinsic scatter. However, the parameters responsible for this simplicity have not been identified. Using the Spitzer Photometry \& Accurate Rotation Curves galaxy catalogue, we argue that the radial acceleration relation (RAR) between galaxies' baryonic and total dynamical accelerations is the fundamental $1$-dimensional correlation governing the radial (in-disk) dynamics of late-type galaxies. In particular, we show that the RAR cannot be tightened by the inclusion of any other available galaxy property, that it is the strongest projection of galaxies' radial dynamical parameter space, and that all other statistical radial dynamical correlations stem from the RAR plus the non-dynamical correlations present in our sample. We further provide evidence that the RAR's fundamentality is unique in that the second most significant dynamical relation does not possess any of these features. Our analysis reveals the root cause of the correlations present in galaxies' radial dynamics: they are nothing but facets of the RAR. These results have important ramifications for galaxy formation theory because they imply that to explain statistically late-type galaxy dynamics within the disk it is necessary and sufficient to explain the RAR and lack of any significant, partially independent correlation. While simple in some modified dynamics models, this poses a challenge to standard cosmology.
\end{abstract}

\begin{keywords}
    galaxies: kinematics and dynamics -- dark matter -- methods: data analysis
\end{keywords}


\section{Introduction}\label{sec:intro}

A key task of astrophysics is to test the galaxy formation scenario posited by the concordance \acl{LCDM} model of cosmology (\acs{LCDM};~\citealt{Rees_Ostriker,White_Rees,White_Frenk}). This may be done by studying unusual or otherwise highly constraining individual or small sets of objects, for example satellites in coherent planes~\citep{Pawlowski}, tidal dwarf galaxies~\citep{TDG}, dark matter deficient galaxies~\citep{Dokkum}, massive high-redshift galaxies~\citep{jwst} and large clusters~\citep{elgordo}. Alternatively one may ask whether the galaxy population as a whole is as expected in \ac{LCDM}. A key feature of galaxy formation in \ac{LCDM} is that it is highly stochastic: not only do the properties of galaxies depend on the complex merger and mass accretion histories of the dark matter haloes in which they live, but chance baryonic phenomena such as supernova and feedback from active galactic nuclei can heat and redistribute mass, significantly altering galaxies' photometry and kinematics. One way to test the model is, therefore, to ask whether this complexity is manifest in galaxy phenomenology.

Surprisingly, galaxies are observed to be rather simple systems. This is most readily shown by applying the \acl{PCA} (\acs{PCA};~\citealt{pearson_PCA,hotelling_PCA}) to galaxy photometry, which calculates the fraction of the covariance in the data that may be accounted for by selected linearly combinations of fluxes in various bands. On multiple data sets and a wide range of variations of the basic technique (e.g. including derived quantities such as star formation rate, size, gas mass and morphology) it is found that only $1$-$3$ principal components are required to explain the great majority ($\gtrsim 90$ per cent) of the data covariance~\citep{Disney, PCA_1, PCA_2, PCA_3, PCA_4, PCA_5, PCA_6, PCA_7}. Nonlinear dimensionality reduction techniques such as variational auto-encoders and self-organizing maps have also been applied, and the results extended to galaxies' spectra~\citep{VAE, LLE, SOM_1, SOM_2, SOM_3, SOM_4}. These studies again indicate that fewer pieces of information are required to reproduce galaxy properties at high fidelity than would be expected in a model where the entire past history of the galaxy and its halo is important.

These findings have analogues in galaxy kinematics, which may be viewed as the set of correlations between the internal motions of galaxies, tracing the underlying potential, and their baryonic mass distributions. To first order the radial kinematics can be summarized by a characteristic velocity and the mass distribution by a total mass and size, leading to the \acl{TFR} (\acs{TFR};~\citealt{Tully-Fisher}) for late-type galaxies and the \acl{FP} (\acs{FP};~\citealt{FP_1, FP_2}) for early-types. The baryonic \ac{TFR} has very small intrinsic scatter and is a near-perfect power-law over five decades of mass~\citep{McGaugh_BTFR, McGaugh_gas}, with few clear outliers~\citep{Pina_1,Pina_2}. This essentially indicates that two independent degrees of freedom suffice to explain the data (the normalization and power-law index), and perhaps more significantly that at fixed mass surface brightness is asymptotically irrelevant for the rotation velocity~\citep{Lelli_smallscatter, Desmond_uncorrelated}. The \ac{FP} is a planar manifold in \{mass, velocity, size\} space, also with very small intrinsic scatter and a tilt relative to the expectation from the Newtonian virial theorem~\citep{Pizagno, Cappellari, Desmond_FJFP}.

In late-type galaxies, the global radial (i.e. in-disk) kinematic behaviour is subsumed by a local relation between the baryonic ($\gbar$) and dynamical ($\gobs$) centripetal acceleration across \acp{RC} known as the mass discrepancy--acceleration or \acl{RAR} (\acs{RAR};~\citealt{Milgrom_1, Sanders_1990,McGaugh_2004,RAR}). This is highly regular and possesses little intrinsic scatter, and may be fit near-perfectly by functions with only $1$-$2$ free parameters~\citep{RAR,Desmond_2023_ESRRAR}. It has recently been shown that the intrinsic or underlying \ac{RAR}---obtained by marginalizing over all relevant galaxy variables---has a scatter of $0.034 \pm 0.002~\mathrm{dex}$ around a $1$-parameter function in the \acl{SPARC} (\acs{SPARC};~\citealt{SPARC}) sample, making it the tightest known dynamical galaxy scaling relation~\citep{Desmond_uRAR}. Evidence is growing that early-type galaxies in fact follow the same \ac{RAR} as late-types~\citep{RAR,Shelest_2020,ellipticals_2, ellipticals_1, ellipticals_3}.

Without a clear prediction or explanation of galaxies' simplicity within concordance cosmology it is important to study it empirically in an attempt to learn the features that govern it. These are the features a theory must account for to explain galaxy phenomenology, and may also provide new empirical tools for measuring galaxies' properties (as the \ac{TFR} and \ac{FP} afford redshift-independent distance probes). Motivated by the striking properties of the \ac{RAR}, we test here the hypothesis that it is the ``fundamental'' $1$-dimensional relation in (late-type) radial galaxy dynamics, meaning that it alone accounts statistically for \emph{all} correlations involving dynamics within the disk. If this is so, then the full set of such correlations amount to nothing more than the \ac{RAR} in conjunction with the non-dynamical correlations present in the sample, and explaining them is tantamount to explaining the \ac{RAR} (and lack of any competing correlation). We break this question into two which we tackle in turn:
\begin{enumerate}
    \item Do the residuals of the \ac{RAR} have no statistically significant correlation with any other galaxy or environmental property? (If they did, the \ac{RAR} is not fundamental because it can be augmented by inclusion of those properties.)
    \item Is the \ac{RAR} the tightest projection of galaxies' radial dynamical parameter space, and can all other projections be explained by the \ac{RAR} in conjunction with non-dynamical correlations present among the galaxy properties? (If not, there is at least one other relation as important as the \ac{RAR}.)
\end{enumerate}
We consider an affirmative answer to both questions to be necessary and sufficient to establish the fundamentality of the \ac{RAR}. We then further assess whether fundamentality is a \emph{unique} property of the \ac{RAR}, or whether there are any other (partially) independent relations also satisfying these fundamentality conditions. These aims are achieved by means of non-linear machine learning algorithms.

The structure of the paper is as follows. In~\Cref{sec:data} we describe the \ac{SPARC} galaxy sample on which we base our analysis. \Cref{sec:method} lays out our methods, including the statistics and methods we employ and the generation of mock data to validate our procedure. \Cref{sec:results} contains our results---separately for each question above---while~\Cref{sec:disc} discusses the broader ramifications of our results, caveats and systematic uncertainties, and useful further work. \Cref{sec:conc} concludes. The reader without particular interest in the technicalities of the methods is advised to skip~\Cref{sec:method} and refer back to it as and when it is mentioned in~\Cref{sec:results}. For brevity, we will henceforth refer to statistically averaged, equilibrium radial motions within the plane of the disk simply as ``dynamics''; these caveats are discussed further in~\Cref{sec:scope}. All logarithms are base-$10$.


\section{Observational data}\label{sec:data}

We utilize the \ac{SPARC} database\footnote{\url{http://astroweb.cwru.edu/SPARC/}} containing $175$ galaxy \acp{RC} with \textit{Spitzer} $3.6~\mu\mathrm{m}$ photometry~\citep{SPARC}. Following~\cite{RAR}, we exclude galaxies with a quality flag of $3$ (large asymmetry, non-circular motion, large offset between stellar and HI distribution), inclination less than $30$ degrees, or fractional uncertainty on the observed rotational velocity exceeding $10$ per cent. We are left with $2,696$ \ac{RC} samples from $147$ late-type galaxies. We define the total disk and bulge surface brightness as $\Sigma_{\rm tot} = \Sigma_{\rm disk} + \Sigma_{\rm bulge}$. Unresolved disk and bulge surface brightness, $\Sigma_{\rm disk}$ and $\Sigma_{\rm bulge}$, are replaced with a single value below the minimum resolved value of $0.01~L_\odot / \mathrm{pc}^2$ (which has no effect on our results).

\cref{tab:features} lists the parameters used for predicting galactic dynamical variables and searching for residual correlations. Local parameters (one value per \ac{RC} point per galaxy) are listed above the line, while galaxy-wide parameters (one value per galaxy) are listed below it. We take the logarithm of properties with a dynamical range spanning several orders of magnitude, as indicated in the final column.

Each of these parameters is contained in the \ac{SPARC} database except for the gravitational field at the \ac{SPARC} galaxies' position, sourced by surrounding baryonic mass. This is calculated in~\cite{Chae_EFE2} and will allow us to assess whether $\gobs$ depends on the external field. At fixed internal mass distribution this would constitute a violation of the strong equivalence principle, as has been argued for by~\citet{Chae_EFE1, Chae_EFE2, Chae_EFE3}. We adopt the results assuming maximal clustering of unseen baryons with visible structures, which maximizes data agreement using the fitting function of~\citet{Chae2022}, as shown in~\citet{Chae_EFE2, Chae_EFE3}. The results assuming no clustering of unseen baryons are however very similar. The values of $\e$ are available for only $90$ galaxies in the \acl{SDSS} (\acs{SDSS};~\citealt{SDSS}) footprint; when including $\e$ as a feature, we fill the missing values with the median of the measured sample.

\begin{table}
    \centering
    \begin{tabular}{lcc}
        Label & Property & Logarithm? \\
        \hline
        \hline
        $\gbar$ & Baryonic centripetal acceleration & \checkmark \\
        $r / R_{\rm eff}$ & Radial distance normalized to $R_{\rm eff}$ & \checkmark \\
        $\Sigma_{\rm disk}$ & Disk surface brightness & \checkmark \\
        $\Sigma_{\rm bulge}$ & Bulge surface brightness & \checkmark \\
        $\Sigma_{\rm tot}$ & Total surface brightness & \checkmark \\
        \hline
        $D$ & Galaxy distance &  \\
        $i$ & Galaxy inclination &  \\
        $L_{3.6}$ & Galaxy luminosity at $3.6~\mathrm{\mu m}$ & \checkmark \\
        $R_{\rm eff}$ & Galaxy effective radius at $3.6~\mathrm{\mu m}$ & \checkmark \\
        $M_{\rm HI}$ & Galaxy neutral hydrogen mass & \checkmark \\
        $T$ & Galaxy type & \\
        $\e$ & External Newtonian field of baryons  & \checkmark \\
        \hline
    \end{tabular}
    \caption{Summary of features considered as potential predictors of galactic dynamical variables. A logarithmic transformation is applied to the indicated features with a wide dynamical range before inputting them into regression models.}
    \label{tab:features}
\end{table}

We calculate the uncertainty of $\gbar$ and $\gobs$ as
\begin{subequations}
\begin{align}
    \delta \gbar &= \frac{1}{10} \frac{\Upsilon_{\rm gas} V_{\rm gas}^2}{r} + \frac{1}{4}\frac{\left(\Upsilon_{\rm disk} V_{\rm disk}^2 + \Upsilon_{\rm bulge} V_{\rm bulge}^2\right)}{r},\\
    \delta \gobs &= \gobs \sqrt{4\frac{\delta \Vobs^2}{\Vobs^2} + 4\frac{\delta i^2}{\tan^2 i} + \frac{\delta D^2}{D^2}},
\end{align}
\end{subequations}
where the numerical factors in $\delta \gbar$ correspond to the mass-to-light ratio uncertainties. We assume no further uncertainty on the velocity components; however, the calculation in~\cite{SPARC} requires a disk thickness which is expected to introduce unmodelled uncertainty to the baryonic velocities at the $\sim 10-15$ per cent level in the innermost region (F. Lelli, priv. comm.). These uncertainties dominate those from other sources such as the covariances across \acp{RC} due to HI interferometry and the integral equations used to calculate $V_{\rm disk}$ and $V_{\rm bulge}$~\citep{SPARC,Lelli2023_MassModels}.
We propagate the uncertainty as $\delta\log \gbar = \delta \gbar / \left(\gbar \ln 10\right)$ and similarly for $\delta \log \gobs$.


\section{Methodology}\label{sec:method}

In this section we detail the statistical tools used in our analysis. We discuss the analytic functions used to approximate the \ac{RAR}, the \ac{PC} analysis, machine learning regression techniques, the loss function, and the generation of \ac{SPARC}-like mock data with controlled dynamics. The main thrust of our analysis, including its results, can be found in~\Cref{sec:results}, which refers back to this section for details.


\subsection{RAR analytic approximations}\label{sec:analytic_funcs}

We consider two functions known to describe the \ac{RAR} well. The first is the \ac{RARIF} introduced by~\cite{RAR}:
\begin{equation}\label{eq:rar_if}
    \gobs = \frac{\gbar}{1 - e^{-\sqrt{\gbar / a_0}}},
\end{equation}
where $a_0 \sim 10^{-10}~\mathrm{m} / \mathrm{s}^2$ is a global characteristic acceleration scale. The second is the \ac{SIFEFE}, which includes an additional parameter $\e$ causing a downturn in the \ac{RAR} at low $\gbar$~\citep{Milgrom_2, Chae2022}:
\begin{flalign}\label{eq:efe_if}
&\gobs = \gbar \left(\frac{1}{2} + \left(\frac{1}{4} + \left(\left(\frac{\gbar}{a_0}\right)^2 + (1.1 \e)^2\right)^{-\frac{1}{2}}\right)^\frac{1}{2}\right) \times \nonumber \\
&\Biggl(1 + \tanh\left(\frac{1.1 \e}{\gbar/a_0}\right)^{1.2} \times \left(-\frac{1}{3}\right) \times \\
&\frac{\left(\left(\left(\frac{\gbar}{a_0}\right)^2 + (1.1 \e)^2\right)^{-\frac{1}{2}}\right) \left(\frac{1}{4} + \left(\left(\frac{\gbar}{a_0}\right)^2 + (1.1 \e)^2\right)^{-\frac{1}{2}}\right)^{-\frac{1}{2}}}{1 + \left(\frac{1}{2} + 2\left(\left(\frac{\gbar}{a_0}\right)^2 + (1.1 \e)^2\right)^{-\frac{1}{2}}\right)^\frac{1}{2}}\Biggr). \nonumber
\end{flalign}
$\e$ may be considered as either a global constant or a galaxy-specific parameter, leading to a slightly distinct \ac{RAR} for each galaxy. Studies of the \ac{SPARC} data have provided evidence for the effect of $\e$~\cite{Chae_EFE1, Chae_EFE2, Chae_EFE3, Chae2022}.

These empirical fitting functions hold particular significance in the theory of \acl{MOND} (\acs{MOND}; \citealt{Milgrom_1, Milgrom_2, Milgrom_3}), an alternative to the Newtonian gravity on which General Relativity and hence \ac{LCDM} is based. \cref{eq:rar_if} is a simple instantiation of the fundamental \ac{MOND} postulate that $\gobs = \gbar$ for $\gbar \gg a_0$ and $\gobs = \sqrt{\gbar a_0}$ for $\gbar \ll a_0$. \cref{eq:efe_if} models the \ac{EFE} in the Aquadratic Lagrangian (AQUAL;~\citealt{Bekenstein1984_AQUAL}) theory: due to the nonlinear acceleration-based modification, galaxies with identical mass distributions but differing environments will have different internal dynamics. A stronger external field pushes a galaxy closer to the Newtonian limit and hence produces a fall-off in the \ac{RAR} at low $\gbar$. In this interpretation, $\e \equiv \gext / a_0$, where $\gext$ represents the strength of the Newtonian gravitational field sourced by the surrounding baryonic mass.


\subsection{Partial correlation}\label{sec:partial_correlation}

We employ a one-sided \ac{PC} analysis~\citep{Baba_PC2004} to investigate the correlation between $\gobs$ and other galaxy properties $\left\{x_k\right\}$ at fixed $\gbar$. We consider the \ac{RAR} in logarithmic space and define the residuals $\Delta \log \gobs \equiv \log \gobs - \mathcal{F}(\log \gbar)$, where $\mathcal{F}$ is a \ac{RAR}-like function that predicts $\log \gobs$ based on $\log \gbar$. We then compute the $\gbar$-controlled \ac{PC} between $\gobs$ and $x_k$ as the Kendall rank correlation $\tau \in \left[-1, 1\right]$ of $\Delta \log \gobs$ and $x_k$, which we denote $\tau(\gobs, x_k | \gbar)$~\citep{Kendall1938}. This choice of correlation coefficient is motivated by the lack of prior expectation for linear residual correlations and by the greater robustness of the Kendall than Spearman coefficient~\citep{kendall_1,kendall_2}. The Spearman coefficient yields very similar results.

The \ac{PC} coefficient does not account for uncertainties, so a non-zero $\tau(\gobs, x_k | \gbar)$ does not necessarily indicate a significant relationship between $\Delta \log \gobs$ and $x_k$. While the correlation coefficient is sensitive to the mean trend, the predictive relationship between specific data realizations might be dominated by noise and the standard $p$-values therefore unreliable. Instead, we assess the impact of uncertainties on the \ac{PC} by generating mock data using the \ac{SIFEFE} with the \ac{SPARC} error model (\Cref{sec:mock}) and calculate the corresponding \ac{PC} coefficients to compare to those of the real data. This yields the distribution of \ac{PC} coefficients corresponding to a \ac{SPARC}-like data set that explicitly contains no dynamical correlations besides the \ac{RAR}. If the data \ac{PC} coefficient is a typical draw from the mock distribution, \ac{SPARC} is a typical realization of a galaxy population in which dynamics is set entirely by the \ac{RAR}.


\subsection{Design of machine learning regressors}\label{sec:regressors}

We investigate the relationship between baryonic and dynamical properties using a fully connected, feed-forward \ac{NN} and decision tree-based regression techniques. The baryonic features considered as potential predictors of a target dynamical quantity are presented in~\cref{tab:features}. For galaxy-wide properties, each \ac{RC} sample is assigned its galaxy's value. All properties are scaled to be zero-centred with a variance of unity.


\subsubsection{Neural network architecture}\label{sec:nn_architecture}

A \ac{NN} represents a continuous, differentiable mapping: $f: \mathbb{R}^m \rightarrow \mathbb{R}^n$ with $m$ input and $n$ output nodes. The simplest fully connected, feed-forward \ac{NN} without hidden layers relates an $m$-dimensional input feature vector to an $n$-dimensional prediction by passing the sum of the product of the feature vector with a weight matrix and a constant bias factor through a non-linear activation function (e.g., sigmoid, hyperbolic tangent;~\citealt{bishop2007,Roberts2021}). The weights and biases are trainable parameters optimized by minimizing a loss function on training samples through back-propagation (\citealt{Rumelhart_backprop}; see~\Cref{sec:fitting}). In \acp{NN} with hidden layers, the output of a previous layer serves as input for the subsequent layer.

Our chosen \ac{NN} architecture is as follows. The network's input layer is normalizing: it adjusts the samples' mean to zero and variance to unity. The network comprises two parallel components with outputs that are combined: a fully-connected \ac{NN} with a wide hidden layer, and a single-node hidden layer with no activation, functioning as a linear operation. For the wide hidden layer, we employ the \ac{LRELU} activation function.
\begin{equation}
    \phi(x) =
    \begin{cases}
    x~&\mathrm{if}~x \geq 0\\
    \alpha x~&\mathrm{if}~x < 0,
    \end{cases}
\end{equation}
where we choose $\alpha = -0.2$. In contrast to the conventional ReLU ($\alpha = 0$), the \ac{LRELU} has a non-vanishing gradient if $x < 0$. We add a dropout layer following the wide hidden layer, which randomly sets a fraction $f$ of nodes' inputs to $0$ and scales the remaining inputs by $1 / (1 - f)$~\citep{Gal2015}. Additionally, we apply an $l^2$ norm to regularize the weights with $\lambda = 0.01$. The loss function in~\cref{eq:loss}, which depends on the derivative of the \ac{NN} prediction with respect to its input, is optimized using the Adam stochastic optimizer~\citep{Adam_Kingma} with default hyperparameters\footnote{\url{https://www.tensorflow.org/api_docs/python/tf/keras/optimizers/Adam}} and a cosine decay learning schedule~\citep{Loshchilov2016}. We implement the \ac{NN} architecture in \texttt{TensorFlow}\footnote{\url{https://www.tensorflow.org}}~\citep{tensorflow2015-whitepaper} and optimize the width of the hidden layer and the dropout rate (\Cref{sec:hyperparameters}).

Training is performed over the training samples containing $60$ per cent of galaxies. This fraction is chosen by fitting the \ac{RAR} as a function of it and optimizing the resulting goodness-of-fit to maximize the models' generalization to unseen observations. We verify robustness of our conclusions to this choice. The train-test split is performed at the galaxy level, rather than on separate \ac{RC} samples, to prevent information leakage between the sets otherwise introduced by the galaxy-wide properties. We choose a batch fractional size of $1/3$, with one epoch consisting of a single pass through all three batches. The training process continues until the validation loss has not decreased for at least $2,000$ epochs, after which the optimal set of weights is restored. Due to the limited sample size, the validation loss is evaluated on the entire training data set.


\subsubsection{Decision tree models}\label{sec:decision_trees}

Decision tree regressors are grown by recursively partitioning samples into increasingly homogeneous subsamples minimizing a specific loss function, such as the weighted variance of the descendant nodes' samples. Predictions are acquired by traversing a path within the tree and procuring the mean value of a terminal leaf. The \acl{RF} (\acs{RF};~\citealt{Breiman_2001_RandomForest}) is a widely employed model. In it, an ensemble of decision trees is constructed, with each tree trained on bootstrap aggregated resamples of the training data and a randomly selected number of features for optimal split determination at each node. This approach mitigates the over-fitting issue commonly encountered in individual decision trees. Another ensemble model is the \acl{ET} (\acs{ET};~\citealt{Geurts_2006_ExtraTrees}). This neither performs bootstrap resampling nor seeks an optimal split, but instead randomly selects a subset of features at each node, assigns them a random split, and divides the samples based on the feature whose split minimizes the loss function. An alternative to constructing large ensembles of independent decision trees is to consider sequential, gradient-boosted trees where each tree is designed to predict the residuals of its predecessors~\citep{Friedman2001, Friedman2002}.

At first glance, tree-based regressors possess an appealing property: the ability to provide feature importance scores. The feature importance is the average summed impurity decrease (homogeneity increase) across all nodes split along a particular feature, averaged over all trees in the ensemble. An alternative metric not limited to tree-based regressors, permutation importance, is computed by permuting a feature, thereby randomizing its correlation with both the remaining features and the target, and subsequently observing the model's relative predictive loss. Notably, the model is not retrained during this process, and the procedure is repeated multiple times to obtain a reliable estimate. High permutation importance indicates that the model's predictive power diminishes upon feature permutation, while low permutation importance suggests that the feature is either irrelevant or strongly correlated with another feature. However, interpreting the allocation of either feature or permutation importance to individual features can be challenging when features exhibit strong mutual correlations. Explainable machine learning methods, such as Shapley additive explanation (\acs{SHAP};~\citealt{lundberg2017_shap,lundberg2020_shap}) attempt to address this issue, although their interpretation is not always straightforward~\citep{krishna2022_disagreement}. Because of this, we do not put strong emphasis on the explainability aspect of our methods and will only briefly rely on the permutation importance score in~\Cref{sec:part2}.

We employ the \ac{ET} model \texttt{ExtraTreesRegressor} from \texttt{scikit-learn}\footnote{\url{https://scikit-learn.org/}}~\citep{scikit-learn}, and the \acl{XGB} gradient boosting decision tree model \texttt{XGBRegressor} from the \texttt{XGBoost}\footnote{\url{https://xgboost.readthedocs.io/}} library~(\citealt{Chen2016_XGBoost}, denoted \ac{XGB} hereafter). We do not use the \ac{RF} as we find its behaviour near-identical to the \ac{ET}. We again allocate $60$ per cent of galaxies to the training set and the remainder to the test set.


\subsubsection{Hyperparameter selection}\label{sec:hyperparameters}

The efficacy and accuracy of any \ac{ML} regressor hinges upon the appropriate selection of its hyperparameters, describing the architecture of the algorithm. Striking an optimal balance in the complexity is crucial, as overly simplistic models may fail to capture relevant features of the data while excessively complex ones risk overfitting and hence poor extrapolation or generalization. We employ \texttt{Optuna}\footnote{\url{https://optuna.org}} for hyperparameter optimization, which facilitates efficient exploration of high-dimensional hyperparameter spaces~\citep{Optuna}. The optimal set of hyperparameters is determined via cross-validation, in each case optimizing a loss function described in~\Cref{sec:fitting}. Because the training of a \ac{NN} is more computationally demanding than that of \ac{ET} and \ac{XGB}, we automatically optimize only the hidden layer width and dropout rate.

We conduct $25$ trials using \texttt{Optuna}, with each trial involving $50$ re-trainings of the \ac{NN} on the same data to account for stochasticity. For the \ac{ET} and \ac{XGB} we execute $10,000$ trials without any retraining in each trial. The hyperparameters that we optimize for each model are presented in~\cref{tab:hyperparam_table}, which also includes the best parameters for the prediction of $\gobs$ from $\gbar$ (i.e. the \ac{RAR}).

\begin{table*}
    \centering
    \begin{tabular}{llll}
        & Hyperparameter & Options & Best for \ac{RAR}\\
        \hline
        \hline
        \texttt{\texttt{TensorFlow}} \ac{NN}    & $\mathtt{width}$                   & $\mathbb{N}(8, 128)$             & $16$                  \\
                                                & $\mathtt{dropout\_rate}$           & $ \mathbb{R}(0.001, 0.10)$       & $0.05$                \\
        \hline
        \texttt{ExtraTreesRegressor}            & $\mathtt{n\_estimators}$           & $\mathbb{N}(64, 128)$            & $92$                  \\
                                                & $\mathtt{max\_depth}$              & $\mathbb{N}(2, 16)$              & $8$                   \\
                                                & $\mathtt{min\_samples\_split}$     & $\mathbb{N}(2, 32)$              & $2$                   \\
                                                & $\mathtt{max\_features}$           & $\{\mathrm{sqrt,~log2,~None}\}$  & $\mathrm{sqrt}$       \\
                                                & $\mathtt{min\_impurity\_decrease}$ & $\mathbb{R}(10^{-14}, 0.5)$      & $4.2 \times 10^{-5}$  \\
                                                & $\mathtt{ccp\_alpha}$              & $\mathbb{R}(10^{-14}, 0.5)$      & $4.4 \times 10^{-12}$ \\
                                                & $\mathtt{max\_samples}$            & $\mathbb{R}(0.1, 0.99)$          & $0.94$                \\
        \hline
        \texttt{XGBoostRegressor}               & $\mathtt{n\_estimators}$           & $\mathbb{N}(16, 128)$            & $125$                 \\
                                                & $\mathtt{max\_depth}$              & $\mathbb{N}(2, 8)$               & $4$                   \\
                                                & $\mathtt{booster}$                 & $\{\mathrm{gbtree,\,dart}\}$     & $\mathrm{dart}$       \\
                                                & $\mathtt{learning\_rate}$          & $\mathbb{R}(0.01, 0.99)$         & $0.94$                \\
                                                & $\mathtt{gamma}$                   & $\mathbb{R}(0, 10)$              & $1.6$                 \\
                                                & $\mathtt{min\_child\_weight}$      & $\mathbb{R}(0.5, 2.5)$           & $2.14$                \\
                                                & $\mathtt{subsample}$               & $\mathbb{R}(0.5, 1)$             & $0.56$                \\
        \hline
    \end{tabular}
    \caption{Hyperparameter ranges of the \ac{NN}, \ac{ET} and \ac{XGB} regressors that are optimized with~\texttt{Optuna}~(\cref{sec:regressors}), with $\mathbb{N}$ and $\mathbb{R}$ representing integer and real ranges, respectively. The rightmost column shows the best hyperparameters for predicting $\gobs$ from $\gbar$, i.e. the \ac{RAR} relation.}
    \label{tab:hyperparam_table}
\end{table*}


\subsection{Scoring regressors}\label{sec:fitting}

Fitting a model $f: \mathbb{R} \rightarrow \mathbb{R}$, be it an analytic function or machine learning regressor, requires a loss function to minimize. In machine learning this is typically considered to be the mean square error (or equivalently coefficient of determination $R^2$) between the data and prediction. This however ignores the measurement uncertainties which are crucial for reliable inference. We therefore consider a Gaussian likelihood function allowing for uncertainties in both the independent ($x$) and dependent ($y$) variable. This necessitates introducing latent variables for the true location of the independent variable, $x_{\rm t}$, on which the full likelihood function depends (e.g.~\citealt{Berger1999}). To eliminate these nuisance parameters one can either marginalize over them with a suitable prior or fix them to their maximum likelihood values as a function of the inferred parameters to make a profile likelihood. The profile likelihood is unbiased in parameters describing the shape of the function because its maximum is by construction at the same point in parameter space as the original likelihood's, and the underestimation of uncertainties is unimportant when one considers, as we do, only best-fit values.

The likelihood of observed values $x,\, y$ given $x_{\rm t}$ and the functional fit $f$ is
\begin{equation}
    p(x, y | x_{\rm t}, I)
    \propto
    \exp\left(- \frac{(y - f(x_{\rm t}))^2}{2 \sigma_{y}^2} - \frac{(x - x_{\rm t})^2}{2 \sigma_{x}^2}\right),
\end{equation}
assuming uncorrelated Gaussian distributions for $x$ and $y$ with uncertainties $\sigma_{x}$ and $\sigma_{y}$, respectively, and omitting normalization terms. Here, $I$ represents the free parameters of $f(x)$. The maximum likelihood estimate of the true independent variable is given by
\begin{equation}
    \hat{x}_{\rm t} = \argmax_{x_{\rm t}} p(x, y | x_{\rm t}, I).
\end{equation}

We expand the function $f(x)$ to first order around the observed value $x$ as
\begin{equation}
    f(x_{\rm t}) = f(x) + (x_{\rm t} - x) \left.\dv{f}{x}\right|_{x} + \ldots.
\end{equation}
As the \ac{RAR} is approximately composed of two linear segments in log-space (and other dynamical correlations are similarly near power-laws), the error induced by the neglected higher order terms is negligible. Substituting the above two equations into the joint probability, we obtain the profile likelihood:
\begin{equation}
    p_\text{prof}(x, y | I)
    \propto
    \exp\left(-\frac{1}{2} \frac{(y - f(x))^2}{\sigma_{y}^2 + \sigma_{x}^2 \left.\mathrm{d}f / \mathrm{d} x\right|_{x}^2}\right),
\end{equation}
where $\left.\mathrm{d}f(x) / \mathrm{d}x\right|^{2}_{x_{\rm 0}}$ projects the uncertainty of $x$ along $y$. For a set of statistically independent observations $\left\{x, y, \sigma_{x}, \sigma_{y}\right\}$, the overall likelihood is given by their product. A derivation of the alternative likelihood with $x_{\rm t}$ integrated out using a uniform prior, which performs better in the presence of intrinsic scatter, can be found in~\cite{Desmond_2023_ESRRAR}.

In accordance with \ac{ML} terminology, we introduce a loss function $\mathcal{L}$ as the negative profile log-likelihood
\begin{equation}\label{eq:loss}
    \mathcal{L}(x, y | I)
    \equiv
    \frac{1}{2} \frac{(y - f(x))^2}{\sigma_{y}^2 + \sigma_{x}^2 \left.\mathrm{d}f / \mathrm{d} x\right|_{x}^2},
\end{equation}
which we denote henceforth simply as $\mathcal{L}$. This is equivalent to the common least-squares (or $\chi^2$) loss with a sample weight of $1 / \sigma^2$ where
\begin{equation}\label{eq:loss_denominator}
    \sigma^2
    \equiv
    \sigma_{y}^2 + \sigma_{x}^2 \left.\dv{f(x)}{x}\right|_{x}^2.
\end{equation}

To propagate the uncertainty of the independent variable, it is essential that the regressor $f(x)$ be differentiable. This condition is satisfied for our analytic functions as well as our \ac{NN} thanks to automatic differentiation (e.g.~\citealt{Baydin2015}). Our \ac{ET} and \ac{XGB} models are however non-differentiable and so we cannot directly minimize~\cref{eq:loss}. We therefore approximate the gradient with an analytic expression prior to model fitting and set the inverse of~\cref{eq:loss_denominator} to be the sample weight. For the \ac{RAR}, we use the gradient of the \ac{RARIF}, assuming a best-fit value of $a_0 = 1.118\times10^{-10}~\mathrm{m} / \mathrm{s}^2$ obtained by fitting this function to the full \ac{SPARC} data set using~\cref{eq:loss}. For multi-dimensional cases where $f: \mathbb{R}^m \rightarrow \mathbb{R}$ with $m > 1$, the loss function in~\cref{eq:loss} is extended by summing over the square of the product of gradient and uncertainty in each independent variable. In certain instances we will ignore the uncertainty in the independent variable (i.e. $\sigma_x = 0$), denoting this loss as
\begin{equation}\label{eq:loss_nograd}
    \mathcal{L}_{0} = \frac{1}{2} \frac{(y - f(x))^2}{\sigma_{y}^2}.
\end{equation}
We verify that these choices have little effect on the minimum loss or the parameter values at which it occurs.


\subsection{Mock data}\label{sec:mock}

To quantify sample variance and aid in interpreting our results, we generate mock data that by construction exhibit a specific dynamical correlation only. These differ only in the values of $\Vobs$ and their correlations with other variables, and are otherwise generated according to the \ac{SPARC} error model. The first set of mocks calculates $\Vobs$ through the $\gbar - \gobs$ correlation of the data (i.e. the \ac{RAR}) and the second through the $\Sigma_{\rm tot} - \Jobs$ relation, where $\Jobs$ is the jerk $\Vobs^3/r^2$. The reason for the latter choice is that $\Sigma_{\rm tot} - \Jobs$ will be found in~\cref{sec:part2} to be the second strongest dynamical correlation in \ac{SPARC}; we wish to see whether either of these relations is sufficient to account for the other correlations. We do not model intrinsic scatter in either relation, which (at the level at which it is present in \ac{SPARC}) would simply weaken slightly the strengths of the correlations.

Both mock data sets account for the covariance of $\Vbar$ and $\Vobs$ across a single galaxy's \ac{RC} induced by uncertainties in distance ($D$), inclination ($i$), mass-to-light ratios (gas $\Ug$, disk $\Ud$, bulge $\Ub$), and luminosity ($L_{3.6}$) by directly sampling them for each galaxy from their prior distributions. We assume Gaussian distributions for all galaxies and denote with an overbar the prior mean as taken from \ac{SPARC}:
\begin{subequations}
\begin{align}
    D &\hookleftarrow \mathcal{G}\left(\bar{D}, \delta D \right),\\
    i &\hookleftarrow \mathcal{G}\left(\bar{i}, \delta i \right),\\
    \log \Ug &\hookleftarrow \mathcal{G}\left(\log 1.0, 0.04\right),\\
    \log \Ud &\hookleftarrow \mathcal{G}\left(\log 0.5, 0.10\right),\\
    \log \Ub &\hookleftarrow \mathcal{G}\left(\log 0.7, 0.10\right),\\
    L_{3.6} &\hookleftarrow \mathcal{G}\left(\bar{L}_{3.6}, \delta L_{3.6} \right),
\end{align}
\end{subequations}
where the standard deviation of $\log \Ug$ corresponds to a $10$ per cent uncertainty on $\Ug$.

We calculate the mock baryonic acceleration and surface brightness as
\begin{subequations}
\begin{align}
    \gbar
    &=
    \frac{L_{3.6}}{\bar{L}_{3.6}}
    \left(\sum_{\mathrm{X}}\frac{\Upsilon_{\rm X} V_{\rm X} |V_{\rm X}|}{\bar{r}} \right) + \frac{\Ug V_{\rm gas} |V_{\rm gas}|}{\bar{r}},\\
    \Sigma_{\rm tot}
    &=
    \left(\frac{L_{3.6}}{\bar{L}_{3.6}}\right) \bar{\Sigma}_{\rm tot}.
\end{align}
\end{subequations}
where $\mathrm{X} \in \{ \mathrm{disk},~\mathrm{bulge}\}$. We scale the disk and bulge contributions along with the surface brightness by the observational error on luminosity, which is however very small. $\gbar$ and $\Sigma_{\rm tot}$ are distance independent; in case of the former velocities are proportional to $\sqrt{D}$ and galactocentric separation is proportional to $D$. We generate $100$ mock realizations of the \ac{SPARC} data set and calculate the mean and standard deviation among them.


\subsubsection{Imposing the RAR}

For this mock data set we employ a fitting function $\mathcal{F}$, specifically the \ac{RARIF} of \cref{eq:rar_if} or \ac{SIFEFE} of \cref{eq:efe_if}, to connect the sampled baryonic acceleration to the ``true'' total acceleration $\langle g_{\rm obs}\rangle = \mathcal{F}(g_{\rm bar})$. Considering that $V_{\rm obs} \propto 1 / \sin i$ and $r \propto D$, we adjust the mock total acceleration for the sampled inclination and distance:
\begin{equation}
    g_{\rm obs}
    \hookleftarrow
    \mathcal{G}\left(\langle g_{\rm obs}\rangle \left(\frac{\sin i}{\sin \bar{i}}\right)^2 \left(\frac{D}{\bar{D}}\right), \frac{2 \delta \Vobs}{\Vobs}\right),
\end{equation}
where the standard deviation accounts for the propagated statistical uncertainty of $\Vobs$ to $\gobs$. The sampled galactocentric distance is $r = r_{\mathrm{t}} D / \bar{D}$. This yields a set of mock ``observed'' values $\left\{r, \gbar, \gobs\right\}$ for the galaxy's \ac{RC} generated with $\mathcal{F}$. When using the \ac{SIFEFE} to generate the mock data, we have that $\gobs = \mathcal{F}(\gbar, \e)$, where $\e$ is sampled according to $\log \e \hookleftarrow \mathcal{G}(\log \bar{e}_{\rm N}, \delta\log \e)$ given the results of~\citet{Chae_EFE2}. We set the free parameter $a_0$ in the mock data by minimizing the loss function~\cref{eq:loss} over the entire \ac{SPARC} data set. For the \ac{RARIF}, we obtain $a_0 = 1.118\times 10^{-10}~\mathrm{m}\mathrm{s}^{-2}$, while for the \ac{SIFEFE} we find $a_0 = 1.170\times10^{-10}~\mathrm{m}\mathrm{s}^{-2}$. In the latter case we simultaneously fit a global parameter $\e$ (for which we find $\log \e = -2.088$), although this is not used in the construction of the mock data. The precise values of these parameters make no difference to the results.


\subsubsection{Imposing the $\Sigma_{\rm tot} - \Jobs$ relation}\label{sec:Jobs_mock}

\begin{figure}
    \centering
    \includegraphics[width=\columnwidth]{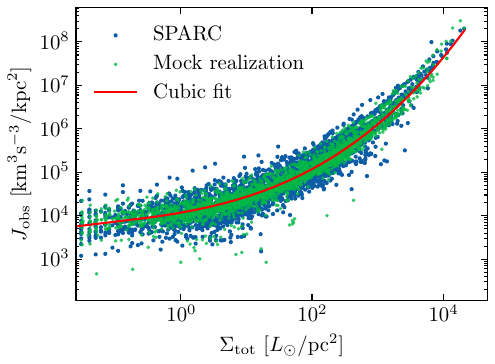}
    \caption{Relation between surface brightness $\Sigma_{\rm tot}$ and jerk $\Jobs = \Vobs^3 / r^2$ for both \ac{SPARC} and an example mock realization. The mock data is generated by assuming a cubic $\Sigma_{\rm tot} - \Jobs$ relation (red line) in conjunction with the \ac{SPARC} error model, as described in the text. This is used as a foil for the \ac{RAR} mock data to establish which dynamical correlations are primary and which derivable from them.}
    \label{fig:mock_sb_vs_jobs}
\end{figure}

To model the $\log \Sigma_{\rm tot} - \log \Jobs$ relation we fit the relation in \ac{SPARC}, shown in~\cref{fig:mock_sb_vs_jobs}, with a cubic polynomial by minimizing~\cref{eq:loss}. This fits the data well and yields coefficients for the cubic, quadratic, linear and constant terms of $4.07,\,0.27,\,0.08,\,0.02$. Our results are not significantly affected by the form of the $\Sigma_{\rm tot} - \Jobs$ function provided it captures the mean trend of the data. We use this to calculate the true mock jerk $\langle J_{\rm obs}\rangle$ from $\Sigma_{\rm tot}$. We then determine the ``observed'' jerk given the inclination and distance for this mock data set, and the statistical uncertainty of $\Vobs$, according to:
\begin{equation}
    \Jobs
    \hookleftarrow
    \mathcal{G}\left(\langle J_{\rm obs} \rangle \left(\frac{\sin i}{\sin \bar{i}}\right)^3 \left(\frac{D}{\bar{D}}\right)^2, \frac{3 \delta \Vobs}{\Vobs}\right).
\end{equation}

\section{Results}\label{sec:results}


\subsection{Residual correlations of the RAR}\label{sec:part1}

Our first objective is to investigate the residual correlations in the \ac{RAR} to determine whether $\gobs$ exhibits correlations with any other variables at fixed $\gbar$. The existence of such correlations would disqualify the \ac{RAR} as fundamental because it would imply that a tighter manifold could be constructed in a higher-dimensional space by incorporating secondary variables. This is analogous to the correlation of velocity dispersion with galaxy size at fixed luminosity in early-type galaxies, which refines the Faber--Jackson relation to the \ac{FP}~\citep{FP_1,FP_2, Bender1992,Cappellari}. To address this, we employ two methods:
\begin{enumerate}
    \item compute partial correlation coefficients between $\gobs$ and all other accessible variables at fixed $\gbar$, and
    \item employ flexible, non-linear machine learning algorithms to determine if the prediction of $\gobs$ can be improved by considering additional variables besides $\gbar$.
\end{enumerate}


\subsubsection{Partial correlation}\label{sec:res_pc}

\begin{figure}
    \centering
    \includegraphics[width=1.\columnwidth]{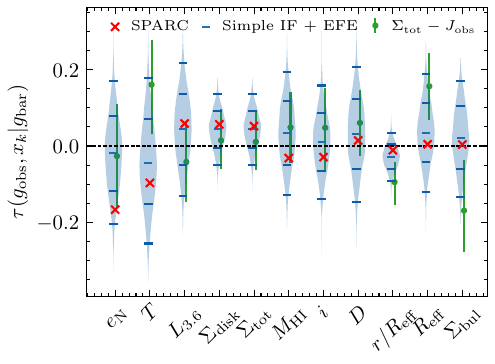}
    \caption{Partial correlation coefficients of the residuals of $\gobs$ at fixed $\gbar$ (as described by the \ac{RARIF}) with various galaxy and environmental features.
    The \ac{SPARC} results are shown by red crosses while those of many mock data sets generated by the \ac{SIFEFE} are shown by blue violins, with the blue ticks indicating the median and $1$ and $2\sigma$ levels. Vertical green bars show $\pm$2$\sigma$ results for mock data with dynamics sourced by the $\Sigma_{\rm tot} - J_{\rm obs}$ relation, with dots at the median. No significant deviations of the data from the \ac{SIFEFE} mocks are observed, indicating that \ac{SPARC} is a typical realization of a population with dynamics specified fully by the \ac{RAR}. The only near-significant secondary correlation of $\gobs$ is with $\e$, hinting at the presence of the \ac{EFE}. The $\Sigma_{\rm tot} - J_{\rm obs}$ mock exhibits non-zero \ac{PC} coefficients incompatible with the data.}
    \label{fig:RARpc}
\end{figure}

We calculate the Kendall correlation coefficient of galaxy properties with the \ac{RAR} residuals around the \ac{RARIF} (\Cref{sec:analytic_funcs}): for the details see~\Cref{sec:partial_correlation}. In~\cref{fig:RARpc} we present the \ac{SPARC} \ac{PC} coefficients as red crosses, and compare them to the expectation from mock data using $\gobs$ generated from the \ac{SIFEFE} (\Cref{sec:mock}), depicted as blue violins. Although several data \ac{PC} coefficients are non-zero, this is consistent with the null hypothesis that the real data is drawn from the mocks. The width of the mock \ac{PC} coefficient distribution is determined by the sensitivity of the respective features to the \ac{RAR} residuals. All violins in~\cref{fig:RARpc} use the same set of resampled residuals. Thus, given the size of the data set and the uncertainties, the \ac{PC} analysis reveals no significant secondary correlations in the \ac{RAR}. We have also tested replacing the \ac{RARIF} fit with our machine learning regressors, finding very similar \ac{PC} values.

We also show mock \ac{PC} coefficients around the \ac{RAR} generated from the $\Sigma_{\rm tot} - \Jobs$ relation as vertical bars extending to $\pm$2$\sigma$ with dots at the median. We find significantly non-zero mock \ac{PC} coefficients in case of $T,\,R_{\rm eff},\,r / R_{\rm eff}$ and $\Sigma_{\rm bul}$, indicating that a different relation between baryonic and dynamical properties would lead to strong residual correlations in the RAR. We further calculate the \ac{PC} coefficients of the $\Sigma_{\rm tot} - \Jobs$ relation of the SPARC data by correlating its residuals around the cubic fit of \cref{sec:Jobs_mock} with baryonic properties (not shown). Comparing these with the expectations from the $\Sigma_{\rm tot} - \Jobs$ mocks, we observe the \ac{SPARC} \ac{PC} coefficients of $T,\,\gbar,\,\Sigma_{\rm bul}$ to be significant at $\geq 3\sigma$. These coefficients in the data are $\tau(\Jobs,\,T|\Sigma_{\rm tot}) = -0.277$, $\tau(\Jobs,\,\gbar|\Sigma_{\rm tot}) = 0.256$ and $\tau(\Jobs,\, \Sigma_{\rm bul}|\Sigma_{\rm tot}) = 0.223$. This shows that the \ac{RAR} is special in not having any residual correlations: it is not simply that such correlations are washed out by measurement uncertainties or the mixing of different galaxy types in \ac{SPARC}.

\begin{figure}
    \centering
    \includegraphics[width=1.\columnwidth]{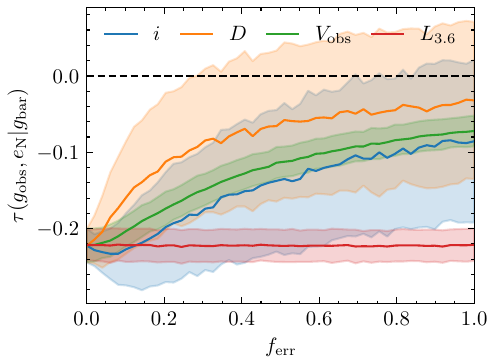}
    \caption{Effect of $i,\,D,\,\Vobs,$ and $L_{3.6}$ uncertainties on the $\gobs$ and $\e$ \ac{PC}, $\tau\left(\gobs, \e | \gbar\right)$. For each parameter, we scale its uncertainty by $f_{\rm err}$ while setting the uncertainty in the remaining parameters to $0$. $f_{\rm err} = 1$ represents the fiducial \ac{SPARC} uncertainty. While in the noiseless case we would expect to see $\tau<-0.2$ due to the presence of the \ac{EFE} in these mocks, the uncertainties in $i,\,D,\,\Vobs$ effectively wash it away. The bands show $1\sigma$ Monte Carlo uncertainty from $1,000$ mock data realizations while the solid lines show the mean.}
    \label{fig:eNpc_scaling}
\end{figure}

For the \ac{RAR}, the most significant partial correlation in \ac{SPARC}, at roughly the 2$\sigma$ level given the RAR mock data, is $\tau(\gobs, \e | \gbar) \approx - 0.2$, suggesting an anti-correlation between $\gobs$ at fixed $\gbar$ and $\e$. This is as expected from the \ac{EFE}. Surprisingly, however, the mock \ac{PC} of $\e$ is approximately centred at $0$ even though it is generated using the \ac{SIFEFE} function with the best-fit (max clustering) values of $\e$ from~\cite{Chae_EFE2}. One would expect a negative \ac{PC} coefficient $\tau(\gobs, \e | \gbar)$ when the mock data includes a non-zero \ac{EFE} and the residuals are calculated using the \ac{RARIF} fit, as increasing $\e$ lowers $\gobs$ at low $\gbar$. However, we show in~\cref{fig:eNpc_scaling} that this effect is washed away by the systematic uncertainties in distance and inclination, as well as the statistical uncertainty in $\Vobs$. We show this by sequentially increasing each uncertainty from $0$ to its \ac{SPARC} value, while keeping the other uncertainties fixed at $0$. Thus, one would not expect to see $\tau(\gobs, \e | \gbar)$ inconsistent with 0 in the real data even if the \ac{EFE} were present at the level suggested by~\cite{Chae_EFE2}.

\subsubsection{Regression with $\gbar$}

We begin our regression analysis by training our machine learning regressors to predict $\gobs$ based solely on $\gbar$ (see~\Cref{sec:regressors,sec:fitting}). This will afford a comparison point when incorporating additional features. We generate $10,000$ test-train splits and follow the procedure outlined in~\Cref{sec:hyperparameters} to determine the optimal hyperparameters for each feature and target.

\begin{figure}
    \centering
    \includegraphics[width=\columnwidth]{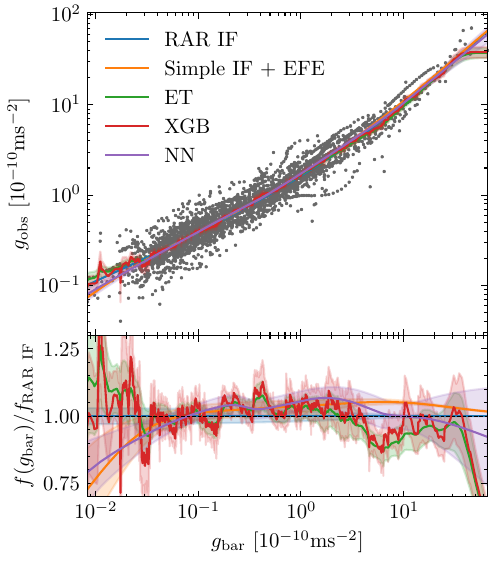}
    \caption{\emph{Upper}: The \ac{SPARC} \ac{RAR} (grey points) with various fits overlaid. The lines show predicted values from analytic functions or \ac{ML} regressors trained on $\gbar$ only, averaged over train-test splits. \emph{Lower}: Ratios of predicted $\gobs$ with respect to the \ac{RARIF}. The bands indicate $1\sigma$ variation between the different test-train splits.}
    \label{fig:RARfit}
\end{figure}

In~\cref{fig:RARfit}, we present the \ac{RAR} fits using the \ac{RARIF}, \ac{SIFEFE}, \ac{NN}, \ac{ET}, and \ac{XGB}. Note that when fitting the \ac{SIFEFE} we use a global value of $\log \e = -2.088$, whereas for the mock data generation using the \ac{SIFEFE} we use the galaxy-by-galaxy best-fit values of~\citet{Chae_EFE2}. Due to the \ac{EFE}, the \ac{RARIF} predicts higher values of $\gobs$ at low $\gbar$ than the \ac{SIFEFE}, whereas elsewhere it is the \ac{SIFEFE} that marginally exceeds the \ac{RARIF}. The regressors successfully recover the mean trend of the \ac{RAR} and are approximately consistent with the analytic functions. The \ac{NN} provides a smooth approximation of the \ac{RAR}, closely aligned with the \ac{SIFEFE}. The \ac{ET} and \ac{XGB} methods struggle to produce a smooth fit due to their discontinuous nature. They also systematically under-predict $\gobs$ at the highest $\gbar$ limit and over-predict it at the lowest $\gbar$ as a result of interpolating between previously seen samples, driving the edge samples closer to the mean.

\begin{figure}
  \centering  \includegraphics[width=\columnwidth]{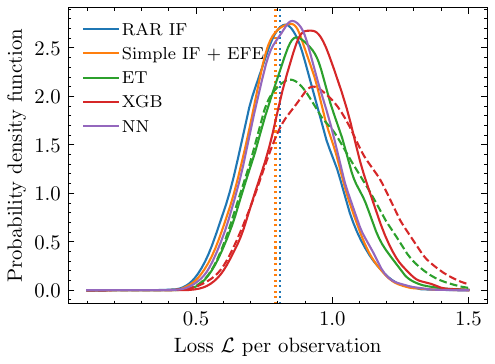}
  \caption{Comparison of \ac{RAR} average test set loss per data point for $10,000$ train-test splits, predicting $\gobs$ from $\gbar$ only (solid lines) and all features (dashed lines). The vertical lines show the \ac{RARIF} and \ac{SIFEFE} results without train-test splitting. The \ac{NN} performs comparably to the analytic functions, while \ac{ET} and \ac{XGB} perform worse. No \ac{ML} regressor improves when given features beyond $\gbar$.}
  \label{fig:RARfitscore}
\end{figure}

In~\cref{fig:RARfitscore} we show the loss $\mathcal{L}$ from~\cref{eq:loss} evaluated on the unseen test samples. Due to the train-test splitting being performed over galaxies rather than the \ac{RC} samples directly, $\mathcal{L}$ is assessed over a varying number of observations. We instead display the mean test set loss per observation. In the absence of train-test splitting (i.e. simply fitting to the entire data set), the \ac{SIFEFE} is favoured over the \ac{RARIF}, as indicated by the vertical lines in~\cref{fig:RARfitscore}. This is practically guaranteed since the \ac{SIFEFE} contains an additional free parameter $\e$, and reduces to the Simple interpolating function, which closely resembles the \ac{RARIF}, when $\e=0$. However, the train-test splitting diminishes this preference, resulting in near-identical distributions of $\mathcal{L}$ for the \ac{SIFEFE} and \ac{RARIF}, with the latter being weakly preferred on average. This suggests that the \ac{SIFEFE} may be slightly over-fitted.

In examining the goodness-of-fit of the \ac{ML} models presented in~\cref{fig:RARfitscore}, we find that the \ac{NN} performs comparably to the \ac{SIFEFE}. The distributions of $\mathcal{L}$ for the \ac{ET} and \ac{XGB} models are systematically higher (i.e. worse) than those of the analytic functions and the \ac{NN}. We therefore find that none of the \ac{ML} models can produce a better description of the data than the \ac{RARIF}; we discuss this point further in~\Cref{sec:cav}.

\subsubsection{Regression with other features}

We now extend our analysis by incorporating additional features alongside $\gbar$ to predict $\gobs$. The dashed lines in \cref{fig:RARfitscore} represent the goodness-of-fit for the \ac{ET}, \ac{XGB}, and \ac{NN} regressors using all available features. Although the peak $\mathcal{L}$ remains relatively unchanged, a more pronounced high-loss tail is observed. Notably, the \ac{NN} regressor exhibits a significantly degraded performance, with a mean loss several times larger, likely due to convergence issues; hence, it is not displayed in \cref{fig:RARfitscore}. We also explore dimensionality reduction via \ac{PCA} for predicting $\gbar$ from all features, allowing \texttt{Optuna} to optimize the number of dimensions. However, this approach yields no improvement.

\begin{figure}
    \centering
    \includegraphics[width=\columnwidth]{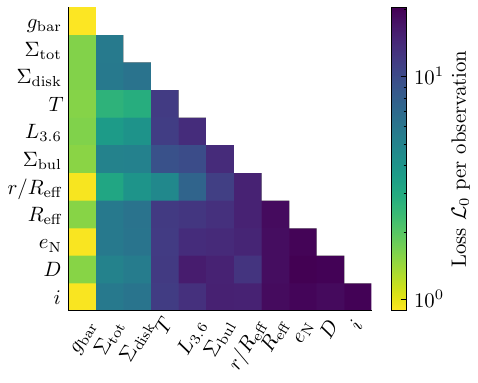}
    \caption{Mean test set loss over $10,000$ test-train splits when predicting $\gobs$ using the \ac{ET} regressor with a single feature (diagonal cells) or pairs of features (off-diagonal cells). The diagonal cells are ordered by increasing loss. $\gbar$ alone is the optimal predictor: its performance is diminished by combining it with any other feature. No other feature, or combination thereof, yields accurate prediction of $\gobs$. The mean and standard deviation of the four lowest loss (yellow) cells are $\gbar: 0.894 \pm 0.152$; $\gbar + i: 0.910 \pm 0.149$; $\gbar + \e: 0.912 \pm 0.152$ and $\gbar + r: 0.920 \pm 0.163$.}
    \label{fig:feature_grid}
\end{figure}

The regressors may be confused when presented with all features, especially if several of them are uninformative. We therefore investigate separately the predictivity of $\gobs$ using all individual and pairs of features. In~\cref{fig:feature_grid}, the diagonal cells display the mean loss $\mathcal{L}_0$ per observation for the \ac{ET} regressor when predicting $\gobs$ using single features. (We verify that other regressors yield similar results.) For simplicity, we present the loss $\mathcal{L}_0$ without propagating the uncertainty in the independent variable, as an approximate value is sufficient to establish the relative merits of the features. We find that $\gbar$ is by far the best individual predictor of $\gobs$. Surface brightness emerges as the next-best predictor, although with a loss nearly an order of magnitude larger. The predictivity of surface brightness can be attributed to its strong correlation with $\gbar$, which is calculated from it. We then assess the predictability of $\gobs$ from pairs of features, shown by the off-diagonal elements of~\cref{fig:feature_grid}. The inclusion of any secondary feature does not enhance the performance of the $\gbar$ predictions, but in fact typically degrades it. All pairs of features that do not include $\gbar$ exhibit poor goodness-of-fit, and it is rarely the case that a pair of features provides more information on $\gobs$ than either one alone. Altogether, similar to the \ac{PC} analysis, these results indicate that no additional information on $\gobs$ can be gleaned from features besides $\gbar$.

\subsection{Projections of the dynamical parameter space}\label{sec:part2}

We have established the \ac{RAR} as the fundamental correlation when $\gobs$ is the target variable, with $\gbar$ fully accounting for $\gobs$ up to apparently random noise. Nevertheless, it remains to be determined whether $\gobs$ is the optimal target variable. There may exist other dynamical variables more strongly correlated with, or predictable from, baryonic properties, which could potentially imply the \ac{RAR} as an approximation. To explore this possibility, we generalize our analysis from~\Cref{sec:part1} to predict an arbitrary dynamical variable, $\mathcal{D}\left(\alpha, \beta\right)$, defined as
\begin{equation}\label{eq:D}
    \mathcal{D}\left(\alpha, \beta\right)
    \equiv \frac{\Vobs^\alpha}{r^\beta}.
\end{equation}

This encompasses as special cases $\Vobs$ ($\alpha=1, \beta=0$), $\gobs$ ($\alpha=2, \beta=1$), jerk $J_{\rm obs}$ ($\alpha=3, \beta=2$) and higher derivatives, without privileging any combination of $\Vobs$ and $r$. The full $2\mathrm{D}$ parameter space of ${\alpha,\beta}$ need not be considered because $\mathcal{D}(\alpha, \beta)^\gamma$ is equally predictable as $\mathcal{D} (\alpha, \beta)$. Without loss of generality we therefore parametrize $\mathcal{D}$ by a new variable $\theta \in [0, \pi)$, defined by
\begin{equation}
    \tan \theta
    \equiv
    \frac{\beta}{\alpha}.
\end{equation}
We consider individual baryonic properties, pairs of properties, and the full set of properties as predictors. The uncertainty in $\Vobs,\,D$ and $i$ is propagated to $\mathcal{D}(\alpha, \beta)$ via
\begin{equation}\label{eq:gencomb_uncertainty}
    \frac{\delta \mathcal{D}(\alpha, \beta)}{\mathcal{D}(\alpha, \beta)}
    =
    \sqrt{\alpha^2 \frac{\delta \Vobs^2}{\Vobs^2} + \alpha^2 \frac{\delta i^2}{\tan^2 i} + \beta^2 \frac{\delta D^2}{D^2}}.
\end{equation}
We generate the mock \ac{RAR}-like data for an arbitrary $\theta$ following the approach of~\Cref{sec:mock} as $\mathcal{D}\left(\alpha, \beta\right) = \gobs^{\alpha / 2} r^{\alpha / 2 - \beta}$, with $\gobs$ computed using the \ac{RARIF} with $a_0 = 1.118\times 10^{-10}~\mathrm{m}\mathrm{s}^{-2}$. This introduces an explicit correlation only between $\gbar$ and $\gobs$. We similarly calculate $\mathcal{D}$ for our mock data by calculating $\Vobs$ from either $\gobs$ or $\Jobs$ and applying \cref{eq:D}. This will let us determine the extent to which \ac{SPARC}-like data with dynamics set solely by the $\gbar-\gobs$ or $\Sigma_\text{tot}-\Jobs$ relation is able to match the predictivity of $\mathcal{D}$ from baryonic variables in the real data.

We use the \ac{ET} and \ac{NN} regressors. For \ac{ET}, we simplify the process by neglecting the propagation of uncertainty in the dependent variable and only generate sample weights using~\cref{eq:gencomb_uncertainty}, showing the loss $\mathcal{L}_0$. The \ac{ET} hyperparameters are fixed to those used in the $\gobs$ regression. For the \ac{NN}, we fully incorporate the uncertainty in the dependent variables in the loss $\mathcal{L}$ given by~\cref{eq:loss} and adopt the \ac{NN} architecture and training described in~\Cref{sec:nn_architecture}.

\begin{figure*}
    \centering
    \includegraphics[width=\textwidth]{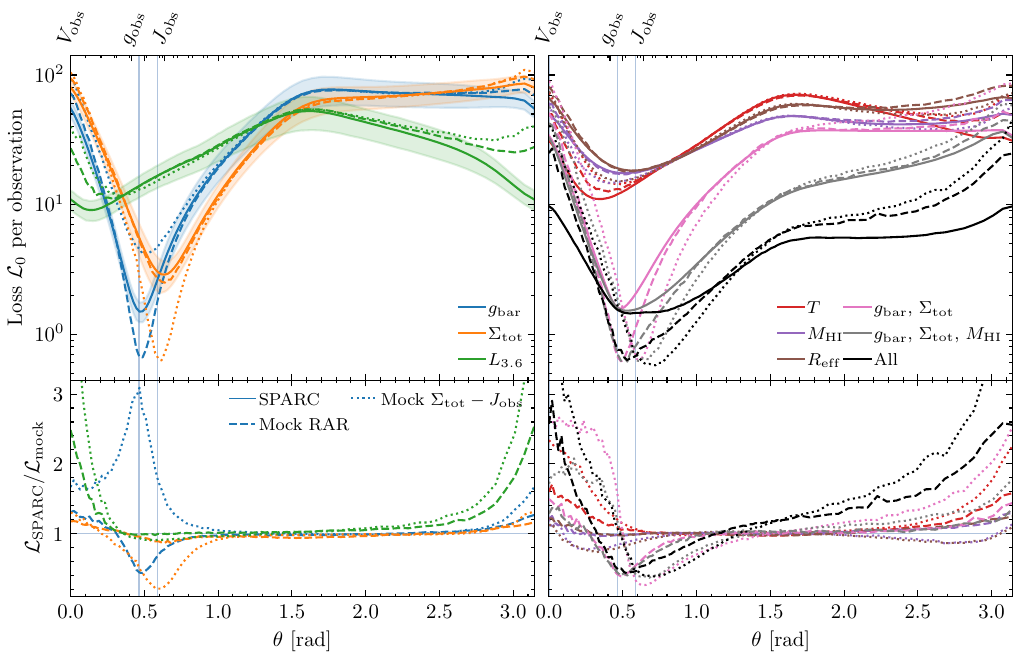}
    \caption{Averaged test set loss $\mathcal{L}_0$ for predicting $\mathcal{D}(\alpha, \beta) = \Vobs^\alpha / r^\beta$ from individual or combinations of features using the \ac{ET} regressor. The generic dynamical variable $\mathcal{D}$ is parametrized by $\tan \theta \equiv \beta / \alpha$. Solid lines show the results in the \ac{SPARC} data set, while dashed and dotted correspond to the mock \ac{RAR} and $\Sigma_{\rm tot} - \Jobs$ data respectively. \emph{Left panels}: $\gbar,\,\Sigma_{\rm tot}$ and $L_{3.6}$ used individually as predictors, with shaded regions indicating $1\sigma$ \ac{SPARC} uncertainty. For clarity, we omit uncertainty bands for the mock data which are similar to those of the real data. \emph{Right panels}: predictivity of other features individually and combinations of features (uncertainty bands omitted). The \emph{lower} panels show the ratios of the \ac{SPARC} and mock-evaluated losses. The \ac{RAR} is the single strongest $1$D relation between galaxy baryonic and dynamical properties, and the mock data using it accounts for the remaining \ac{SPARC} correlations which therefore appear as nothing but aspects of the \ac{RAR}. On the other hand, the mock $\Sigma_{\rm tot} - \Jobs$ mock data fails to explain the \ac{RAR} as well as other observed dynamical correlations.}
    \label{fig:gencomb_ET_double}
\end{figure*}

We present the goodness-of-prediction of $\mathcal{D}(\alpha, \beta)$ by various features according to the \ac{ET} model in~\cref{fig:gencomb_ET_double}. We compute $\mathcal{L}_0$ as a function of $\theta$ for individual features listed in~\cref{tab:features} and for all features combined, showing the results for the real data in solid lines. We focus on the individual dependence of $\gbar,\, \Sigma_{\rm tot},\,L_{3.6}$, as the remaining features are either non-informative or exhibit no interesting trends. We find that $\gbar$ is the best predictor of $\gobs$ ($\theta = \arctan 1/2$), in agreement with the results of~\Cref{sec:part1}. Interestingly, we find that $L_{3.6}$ is the best predictor of $V_{\rm obs}$. This is reminiscent of the \ac{TFR} but surprising because $L_{3.6}$ is a galaxy-wide feature, which cannot therefore capture variations in $\Vobs$ within a single galaxy. $\Sigma_{\rm tot}$ is as good a predictor of $\Jobs = \Vobs^3 / r^2$ as $\gbar$, and $\Jobs$ is approximately the dynamical quantity most predictable from $\Sigma_{\rm tot}$. That the $\Jobs-\Sigma_\text{tot}$ relation is the second strongest correlation in the data is what motivates our decision to investigate mock data created from it, as a foil for the \ac{RAR} results. This is discussed further in~\Cref{sec:imp}.

The inclusion of additional features does not enhance the prediction of $\gobs$, corroborating the results of~\Cref{sec:part1}. Nevertheless, when considering all features simultaneously there is an improvement in predicting $\mathcal{D}$ at $\theta \gtrsim \arctan(1 / 2)$ (black line) where the loss is as low as at $\gobs$. In fact, at the global minimum corresponding to $\tan \theta \approx 0.62$ the loss $\mathcal{L}_0$ per observation is lower by $\sim 0.05$ than the $\gbar-\gobs$ relation (i.e. the \ac{RAR}). However, the $1\sigma$ Monte Carlo uncertainties of the loss are approximately $\sim0.34$ and $0.26$ for the minimum of the black line and the \ac{RAR}, respectively, showing that the improvement is statistically insignificant. We also show the predictability of $\mathcal{D}(\theta)$ using $\gbar,\,\Sigma_{\rm tot}$ and then with $\gbar,\, \Sigma_{\rm tot},\, M_{\rm HI}$. We observe that the addition of $M_{\rm HI}$ significantly reduces the loss, even though it is not informative by itself.

The dashed and dotted lines in~\cref{fig:gencomb_ET_double} show analogous results obtained from the \ac{RAR} and $\Sigma_{\rm tot}- \Jobs$ mock data, respectively (see~\cref{sec:mock}). The mock \ac{RAR} lines are remarkably consistent with the real data, showing that the \ac{RAR} alone is approximately able to account for all dynamical correlations present in \ac{SPARC}. Minor discrepancies arise in the case of predicting $\Vobs$ from $L_{3.6}$ at $\theta \sim 0$ and $\gobs$ from $\gbar$. In the former, a stronger relation is found in the real data compared to the mock data, while in the latter, the relation is stronger in the mock data. These deviations are significant at less than the $\sim$2$\sigma$ level given the formal uncertainties derived from the test-train splitting. However, we do not expect the mock data generation to encapsulate all nuances of real galaxies: the improved prediction at $\gobs$ can be attributed to the absence of intrinsic scatter in the mock \ac{RAR} data. We evaluate the \ac{RMS} deviation of $\log \gobs$ from the \ac{RARIF} fit with respect to both the \ac{SPARC} data and \ac{RARIF} mocks, obtaining values of $\sim 0.133$ and $0.128~\mathrm{dex}$, respectively. In mocks, this originates from accounting for the systematic uncertainties associated with $\Upsilon_{\rm bulge},\,\Upsilon_{\rm disk},\,\Upsilon_{\rm gas},\,i,\,D$ and $L_{3.6}$, as well as statistical uncertainty of $\Vobs$ (see \cref{sec:mock}). If we include a statistically independent intrinsic scatter its best-fit value is therefore $\sim 0.034~\mathrm{dex}$, which agrees perfectly with that estimated by the more sophisticated analysis of \cite{Desmond_uRAR}. Other discrepancies may be due to by minor systematics in the data such as variations in disk thickness, inclination or mass-to-light ratios across the galaxies.

In contrast, the mock $\Sigma_{\rm tot} - \Jobs$ data does not accurately reproduce the dynamical \ac{SPARC} correlations. Most notably, the \ac{RAR} shows a much larger loss when predicting $\mathcal{D}$ from $\gbar$, compared to the mock \ac{RAR} data set which captures the $\Sigma_{\rm tot} - \Jobs$ correlation with a loss comparable to that of \ac{SPARC}. It can also be seen that the $\Sigma_{\rm tot} - \Jobs$ mock has behaviour discrepant with \ac{SPARC} when predicting from $L_{3.6}$. We verify that the agreement between \ac{SPARC} and the \ac{RAR} mock, and disagreement with the $\Sigma_{\rm tot} - \Jobs$ mock, extends also to features not shown in~\cref{fig:gencomb_ET_double}.

\begin{figure}
    \centering
    \includegraphics[width=\columnwidth]{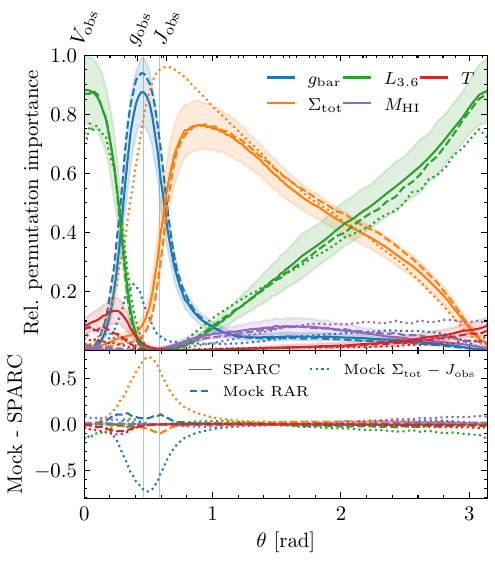}
    \caption{Relative permutation importances of selected features for predicting $\mathcal{D}(\alpha, \beta)$ using the \ac{ET} regressor. Solid lines denote \ac{SPARC} data (with $1\sigma$ uncertainty bands), while the dashed and dotted lines indicate the mock \ac{RAR} and mock $\Sigma_{\rm tot} - \Jobs$ data, respectively. The lower panel shows the difference between the mock and \ac{SPARC} importances. The shown features account for practically all of the model's predictivity at all $\theta$. As in~\cref{fig:gencomb_ET_double}, the consistency between the real and mock data shows that the relative importances of all features derive from the \ac{RAR} in conjunction with the non-dynamical correlations present in \ac{SPARC}.}
    \label{fig:ET_perm}
\end{figure}

We present the corresponding \ac{ET} relative permutation importances of $\gbar,\,\Sigma_{\rm tot},\,L_{3.6},\, M_{\rm HI}$, and $T$ in~\cref{fig:ET_perm}. This quantifies the average increase of $\mathcal{L}_0$ per observation when a feature is permuted, normalized such that the sum of all importances is unity at each $\theta$. These five features account for the majority of the \ac{ET}-inferred importance. In the real data we observe a weak importance of galaxy type around $\Vobs$, while at $\gobs$ the \ac{ET} recognizes $\gbar$ as clearly the most important feature despite its strong correlation with $\Sigma_{\rm tot}$. In case of $\Jobs$, we see the permutation score to be significant for both $\gbar$ and $\Sigma_{\rm tot}$, however from~\cref{fig:gencomb_ET_double} we know that the dependence of $\Jobs$ on $\Sigma_{\rm tot}$ is a consequence of the \ac{RAR} and, therefore, this is again just a case of $\gbar$ and $\Sigma_{\rm tot}$ being strongly correlated. The dashed and dotted lines again show the permutation importances for the mock \ac{RAR} and $\Sigma_{\rm tot} - J_{\rm obs}$ data respectively. While the former are again in agreement with the real data, the importance of $\gbar$ and $\Sigma_{\rm tot}$ in the $\Sigma_{\rm tot} - J_{\rm obs}$ mocks disagrees strongly. This shows that the $\Sigma_{\rm tot} - J_{\rm obs}$ relation is a reflection of the \ac{RAR}, not the other way around. Qualitatively similar conclusions hold for the decision tree importance scores calculated as the average impurity decrease due to a feature.

\begin{figure}
    \includegraphics[width=\columnwidth]{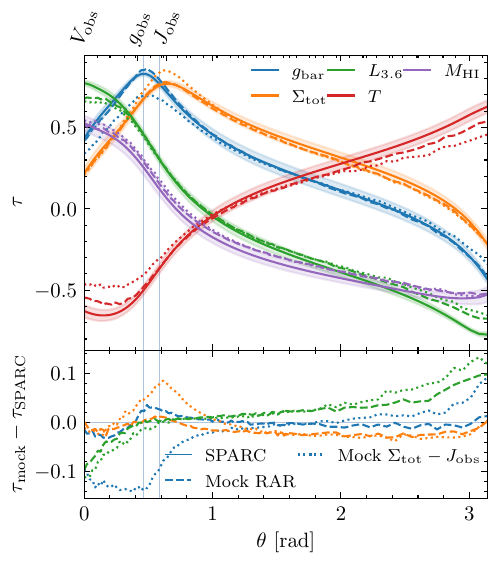}
    \caption{Kendall's $\tau$ correlation coefficient between $\mathcal{D}(\alpha, \beta)$ and $\gbar,\,\Sigma_{\rm tot},\,L_{3.6},\,T,\,M_{\rm HI}$. Solid lines represent the \ac{SPARC} data, dashed lines the mock \ac{RAR} data, dotted lines the mock $\Sigma_{\rm tot} - \Jobs$ data and bands the 1$\sigma$ \ac{SPARC} uncertainties. The lower panel shows the difference between the mock and \ac{SPARC} correlation coefficients, illustrating the superiority of the \ac{RAR} mock data. The results corroborate and reinforce those of~\cref{fig:gencomb_ET_double,fig:ET_perm}.}
    \label{fig:gencomb_kend}
\end{figure}

We repeat this analysis with the \ac{NN} regressor, optimizing the loss $\mathcal{L}$. We focus only on the most important features---$\gbar$, $\Sigma_{\rm tot}$, and $L_{3.6}$---due to the increased computational cost associated with training the \ac{NN}. This plot is not shown because despite the use of a different loss function, we observe almost identical trends to the \ac{ET} analysis (\cref{fig:gencomb_ET_double}). This shows that the results are not sensitive to the regressor used. Finally, to crosscheck the ML models' findings we calculate the Kendall correlation coefficient of $\mathcal{D}(\alpha, \beta)$ with $\gbar,\,\Sigma_{\rm tot},\,L_{3.6},\,T,\,M_{\rm HI}$ and $R_{\rm eff}$~\cref{fig:gencomb_kend}. We again find that the strongest dynamical-to-baryonic correlation is between $\gbar$ and $\gobs$, with strong consistency between the real and \ac{RAR} mock data. Again as in the \ac{ET} analysis, the mock $\Sigma_{\rm tot} - \Jobs$ data fails to explain the dynamical correlations of \ac{SPARC}: it predicts too strong a $\Sigma_{\rm tot} - \Jobs$ relation, and too weak a $\gobs-\gbar$, $L_{3.6}-\Vobs$ and $T-\Vobs$ relation. Scatter in the mock data or additional noise or systematics will only weaken these trends further in the mock data. By contrast, the mock \ac{RAR} data recovers the $\Sigma_{\rm tot} - \Jobs$ relation (indeed the full $\Sigma_{\rm tot} - \mathcal{D}$ relation) near-perfectly.

\section{Discussion}\label{sec:disc}


\subsection{Implications of results}\label{sec:imp}

We demonstrate that in the \ac{SPARC} sample the \ac{RAR} is both necessary and sufficient to account for the full set of correlations among galaxy properties involving dynamical variables, so that it may be considered the fundamental relation of late-type galaxy dynamics. The fact that our \ac{ML} regressors are unable to produce superior fits to the $\gbar - \gobs$ relation than the simple analytic expressions further testifies to the relative simplicity of the \ac{SPARC} \ac{RAR}, the information in which is almost fully captured by simple $1$-$2$ parameter functions. Consequently, the primary task in comprehending these galaxies' dynamics lies in understanding the emergence of the \ac{RAR} and non-emergence of any alternative or additional correlation.

While the \ac{RAR} is an immediate consequence of \ac{MOND}~\citep{Milgrom_1}, its status in a \ac{LCDM} cosmology is debated. Several authors argue that aspects of the \ac{RAR} are highly challenging to reproduce in a \ac{LCDM}-based galaxy formation scenario~\citep{Wu_Kroupa,Desmond_MDAR,Tenneti,McGaugh_tale}, while others argue that it emerges more or less naturally in hydrodynamical simulations or semi-analytic models~\citep{vdB,Keller,Ludlow}. Even if the latter were so, the discovery that the \ac{RAR} is apparently \emph{the only} fundamental correlation of late-type galaxy dynamics provides an additional hurdle for such models. This adds to the existing challenges faced by $\Lambda\mathrm{CDM}$ models attempting to explain the \ac{RAR}~\citep{DC_Lelli,Desmond_MDAR,Keller,Ludlow,Paranjape_Sheth,Desmond_uRAR}.

In~\Cref{sec:part2}, we showed that $L_{3.6}$ is individually the best predictor of $\Vobs$, as expected from the \ac{TFR}~\citep{Tully-Fisher,McGaugh_BTFR}. This result was replicated using mock \ac{RAR} data, indicating that the \ac{TFR}-like $\Vobs - L_{3.6}$ relation can be derived from the \ac{RAR}. It is important to note however that $\Vobs$ is a local quantity, varying across a single galaxy's \ac{RC}, while the \ac{TFR} is defined in terms of an overall characteristic velocity. The correlation is strongest when using the flat rotation velocity~\citep{Ana_2,Ana_1,Lelli_Vs}, which is only captured in the final $\Vobs$ values of some \ac{SPARC} galaxies. It is therefore unsurprising that the $\Vobs - L_{3.6}$ relation we find is significantly weaker than the true \ac{TFR}. Additionally, we found that $\Jobs$ is well predictable individually by both $\gbar$ and $\Sigma_{\rm tot}$ (which are strongly correlated as the former is calculated from the latter), with the $\Sigma_{\rm tot} - \Jobs$ correlation being a reflection of the \ac{RAR}. We demonstrated that the $\Sigma_{\rm tot} - \Jobs$ correlation alone is insufficient to replicate the dynamical correlations in \ac{SPARC}: it fails to accurately reproduce the observed \ac{RAR}, while the \ac{RAR} completely explains the observed $\Sigma_{\rm tot} - \Jobs$ relation. Additionally, its residuals are correlated with baryonic properties in the data, and if it were the correlation driving the dynamics it would induces residual correlations in the \ac{RAR}, which are not observed.

We hypothesize that the $\Sigma_{\rm tot} - J_{\rm obs}$ correlation is caused by the specific disk matter distribution of \ac{SPARC} galaxies, in conjunction with the \ac{RAR} determining the dynamics. To test this, we calculate $\gbar$ for model galaxies using (1) a thin exponential density profile~\citep[Eq. 11.30]{MBW}, and (2) the Kuzmin disk (\citealt{kuzmin};~\citealt{Binney}, Eq. 2.68a) for the baryonic mass distribution. We then calculate $\gobs$ and hence $\mathcal{D}(\theta)$ from the \ac{RARIF}. In both cases, we observe a maximum correlation between $\mathcal{D}(\theta)$ and surface brightness around $\theta=\arctan(2/3)$, i.e. $\Jobs$, as in the real data, although the correlation of the disk solution peaks closer to $\theta=\arctan(1/2)$, i.e. $\gobs$. This shows that the fact that the $\Sigma_\text{tot}-\mathcal{D}$ relation peaks almost exactly at $\Jobs$ in \ac{SPARC} is somewhat coincidental because it depends on the baryonic mass profile; for a different sample the extrema of $\mathcal{L}_0$ and $\tau$ may occur at a $\theta$ different by some tenths of a radian.

We assumed that both the \ac{RAR} and $\Sigma_{\rm tot} - \Jobs$ relations have no intrinsic scatter when generating the mock data. While this assumption is approximately true for the \ac{RAR} (\citealt{Desmond_uRAR} measured it to be $0.034 \pm 0.002~\mathrm{dex}$), it is unlikely to hold for $\Sigma_{\rm tot} - \Jobs$. However, adding uncorrelated intrinsic scatter to the latter relation can only weaken the correlations, deteriorating the prediction of the \ac{RAR} still further. While it is possible to introduce correlations of the residuals with other variables and fine-tune them to reproduce the correlations seen in the data, this will simply approximate the success of the mock data generated solely according to the \ac{RAR}. Thus the \ac{RAR} satisfies the fundamentality criterion of~\Cref{sec:intro} that it can explain the other dynamical correlations, while the $\Sigma_{\rm tot} - \Jobs$ relation does not. Since that relation is the second strongest $1$-dimensional correlation in the data (measured by $\mathcal{L}_0$, $\mathcal{L}$ or $\tau$), it follows that no dynamical correlation in the \ac{SPARC} data set besides the \ac{RAR} can be fundamental: the rest will make even worse predictions for the \ac{RAR} and $\mathcal{D}(\theta)$ generally.

Within the \ac{MOND} paradigm our analysis is capable of shedding light on the presence or absence of the \ac{EFE}. We confirm previous results~\citep{Haghi_EFE, Chae_EFE1, Chae_EFE2, Chae_EFE3, Desmond_uRAR} that the inclusion of the \ac{EFE} improves the fit to galaxy kinematics (i.e. the \ac{SIFEFE} relative to the \ac{RARIF}), and show that the largest partial correlation of $\gobs$ at fixed $\gbar$ is with the external field strength $\e$ (\cref{fig:RARpc}). This suggests the external field to be the most important quantity besides $\gbar$ in the determination of $\gobs$. However, when performing test-train splits the \ac{SIFEFE} no longer outperforms the \ac{RARIF} (\cref{fig:RARfitscore}), and none of our regressors can utilize $\e$ to improve the prediction of $\gobs$ (shown for the \ac{ET} regressor in~\cref{fig:feature_grid}). We therefore do not find significant evidence for the \ac{EFE}, and conclude that better data is required to establish the validity or breakdown of the strong equivalence principle in galaxy \acp{RC}. This conclusion agrees with~\cite{Paranjape_Sheth2}, who challenge the \ac{EFE} detection of~\cite{Chae_EFE2} on the basis of large \ac{RC} uncertainties and potential inaccuracies in estimating the external field strength. Moreover, they argue that \ac{LCDM} naturally predicts an \ac{EFE}-like correlation, albeit with an opposite sign to that of \ac{MOND}.


\subsection{Caveats and systematic uncertainties}\label{sec:cav}


\subsubsection{Scope of the conclusions}\label{sec:scope}

As mentioned at the end of~\Cref{sec:intro}, it is important to stress that our results on the fundamentality of the \ac{RAR} do not extend to late-type galaxy dynamics \emph{in their totality}, but only to \emph{equilibrium radial} dynamics. The \ac{RAR} is limited to motion in the plane of the disk and does not cover the perpendicular, vertical motion which also possesses much interesting dynamics. Examples include vertical velocity dispersion in disks (e.g.~\citealt{vertical_1,vertical_2}), vertical ``breathing modes'' (e.g.~\citealt{breathing,Ghosh2022_spiral}) and the more complicated anisotropic motions of bars and bulges. There also exist dynamical scaling relations in late-type galaxies that involve the vertical gravitational field such as the central baryonic--dynamical surface density relation~\citep{CSDR}. The perpendicular structure of the disk is important for phenomena such as the ``Freeman law,'' an observed upper limit to disk surface densities~\citep{Freeman_1,Freeman_2,Freeman_3}, which may arise from stability requirements involving the full 3D structure~\citep{Freeman_MOND_2}. Out-of-equilibrium and non-axisymmetric dynamics cause deviations from the regular radial motions encapsulated by the \ac{RAR} and are therefore also not covered by our analysis.

The other important restriction to the scope of our conclusion, even within the radial dynamics of late-types, is that the \ac{RAR} is not a substitute for the galaxy-by-galaxy study of \acp{RC}. Our conclusions therefore only hold for equilibrium radial dynamics when stacked across many galaxies and viewed statistically, i.e. \emph{on average}. Although the \ac{RAR} contains the correlation of the \acp{RC} with the baryonic mass distribution, particular points may be of special interest without standing out in the \ac{RAR} plane. An example is ``Renzo's rule,'' a point-by-point correlation of features in the baryonic mass profile with features of the \acp{RC}~\citep{Renzo,Famaey_McGaugh}. While obvious in case the baryons dominate the mass budget, this is difficult to understand in \ac{LCDM} if the \ac{RC} is dominated by the dark matter as this is supposed to be featureless and uncorrelated with the baryons. Velocity measurements for which this is pronounced are rare, so a model may predict the statistics of the \ac{RAR} successfully on average while completely missing this behaviour: it would introduce only a few, relatively small residuals. Thus it cannot be said that explaining the \ac{RAR} alone is sufficient to account in full generality for late-type galaxy dynamics, but only for the average equilibrium radial dynamics, parametrized by $\mathcal{D}$, that we have studied here.


\subsubsection{Goodness-of-fit \& model complexity}

\cref{fig:RARfitscore} shows that while our \ac{NN} performs comparably to the \ac{RARIF} and \ac{SIFEFE}, the losses achieved by the \ac{ET} and \ac{XGB} regressors are systematically higher. This is intriguing, given that the \ac{ML} models are nearly arbitrarily complex and hence should be able to fit the data better. The suboptimal performance of the \ac{ML} models could potentially be attributed to three factors: (1) the sampling of the underlying \ac{RAR} distribution in \ac{SPARC}, (2) overfitting and imperfect hyperparameter optimization, and/or (3) edge effects. We discuss each of these in turn.

First, the samples on the $\gbar - \gobs$ plane are not independent and identically distributed, but instead are strongly correlated within each galaxy. Coupled with the relatively small sample size, this may result in inadequate representation of the test set by the training set galaxies, leading to suboptimal performance of the \ac{ET} and \ac{XGB} models. These models can only interpolate discontinuously between past observations, whereas the analytic functions, constrained by their functional form, and \ac{NN}, which smoothly interpolates between past observations, are more robust in this regard. When training and evaluating the \ac{ML} regressors on all data (i.e., without a test-train split), they ``outperform'' the analytic functions due to overfitting.

Second, as depicted in the lower panel of~\cref{fig:RARfit}, the \ac{XGB} and \ac{ET} models generate a markedly jagged fit, suggesting potential overfitting on training samples. Enhancing the generalization capability of \ac{ML} models often involves hyperparameters optimization, a crucial aspect of our analysis. We employed an automatic hyperparameter optimization using \texttt{Optuna}, conducting the optimization on the full training data until convergence and selecting hyperparameter that minimize the loss on validation samples selected via cross-validation. The choice of the number of folds is crucial as it determines the balance between training and validation; both extremes may lead to a non-representative validation loss. We opted for $5$-fold cross-validation to optimize the hyperparameters, although the goodness-of-fit was similar when using either $3$- or $4$-fold cross-validation instead. The prime method of reducing overfitting is to reduce the complexity of the regressors, for example by restricting the maximum depth of decision trees or the minimum number of samples required to split a node. Although these are already optimized by \texttt{Optuna} (see~\cref{tab:hyperparam_table}), it is possible that a further small improvement could be achieved by hand-tuning them. Indeed the complexity threshold before overfitting occurs appears to be very low on the \ac{RAR} data: as shown in~\cref{fig:RARfitscore}, the \ac{SIFEFE} fits the entire \ac{SPARC} data set better than the \ac{RARIF} but worse when train-test splitting is implemented. This function with only two free parameters is therefore likely overfitted.

Third, in~\cref{fig:RARfit} the performance of the \ac{ET} and \ac{XGB} models deteriorates near the boundaries of the data set, where the average loss per point is higher than average. This limitation arises from the decision trees' inability to extrapolate beyond past observations, resulting in predictions biased towards the mean. The impact of this on the loss distribution is, however, on the order of a few per cent, and is therefore not the main cause of the poor performance of the \ac{ET} and \ac{XGB} regressors.


\subsubsection{Neural network convergence}

Although our \ac{NN} allows derivatives to be propagated fully into the loss function and achieves as high accuracy as the \ac{RARIF}, it is not without problems. Despite extensive training, convergence to a global minimum of the loss is in practice not guaranteed due to the high dimensionality of the parameter space of the \ac{NN} weights and biases. Indeed, whether an \ac{NN} has converged to the global minimum is generally unclear. One way to address this is to use an ensemble of \acp{NN}, where the predictions are stacked to quantify the predictive disagreement arising purely from training~\citep{stiskalek2022_scatter}. An alternative approach to enhance the accuracy and generalizability is to incorporate additional information (``physics'') during training to create a ``\acl{PINN}'' (\acs{PINN};~\citealt{karniadakis2021_pinn, cuomo2022_pinn}). This may involve modification to the network architecture, loss function and/or activation functions. This however comes at the cost of reducing the agnosticism and purely empirical nature of the \ac{NN}, and the simple exercise of replacing the linear connection of the input and output layer in our NN with the \ac{RARIF} (optimizing $a_0$ alongside the other NN parameters) produced no improvement relative to the \ac{RARIF} alone. This again supports the conclusion that residuals of the \ac{RAR} around the \ac{RARIF} are devoid of physical information.


\subsubsection{Raw data reduction \& SPARC selection}

The \ac{SPARC} data contains various systematics unaccounted for in our mock data or likelihood model. This includes inclination variations across the disks in the tilted ring fits, potential variations in mass-to-light ratios and disk thickness across the \ac{RC} and other assumptions made about the structure of the baryonic mass when solving for the \ac{RC} such as the vertical distribution of the gas density~\citep{barolo}. However, \cite{Mancera2022_flaring} demonstrated that gas disk flaring is only important for halo properties inferred from RC fitting in the smallest, gas-dominated dwarf galaxies. The data is also selected inhomogeneously, based on the high-quality \acp{RC} that happen to have been derived in the literature. Nevertheless, \ac{SPARC} provides a thorough sampling of the physical parameter space occupied by rotationally supported galaxies, and there is no obvious reason to expect systematics or selection to bias our results. Some suggestions for the more sophisticated modelling required to corroborate our conclusions with greater precision are given below.


\subsection{Suggestions for further work}\label{sec:further}


\subsubsection{Increasing the robustness of the inference}

Our likelihood functions~\cref{eq:loss,eq:loss_nograd} assume that all of the points on the \ac{RAR} are independent. This is not true in detail because the methods used to derive the \ac{RC} of a given galaxy necessarily couple the points, and several of the parameters relevant to $\Vobs$ and $\Vbar$---distance, inclination and mass-to-light ratios---are properties of the galaxies as a whole. It is possible to infer these parameters simultaneously with a fit to the \ac{RAR}~\citep{Desmond_uRAR}, but this assumes a functional form for the fit and hence is not useful for our more agnostic analysis here. It would however be possible to construct a covariance matrix for $\gobs$ and $\gbar$ within each galaxy by Monte Carlo-sampling $D$, $i$ and $\Upsilon_{\rm X}$ and use that in the Gaussian likelihood, thus assuming that galaxies are independent while \ac{RC} points for a given galaxy are not.

A truly irreproachable demonstration of the fundamentality of the \ac{RAR} would require showing that its residuals correlate with \emph{no} other galaxy or environment property. Here we are restricted to features present in the \ac{SPARC} data set plus $\e$ as calculated in~\citet{Chae_EFE2}. Other interesting features could include colour, star formation rate and environmental density (or functions thereof) on a range of scales, which could be acquired by cross-correlating \ac{SPARC} with other surveys and repeating the calculations of~\citet{Chae_EFE2} for other environmental measures. As an example, $\sim2/3$ of the \ac{SPARC} galaxies were observed in the \ac{SDSS} and hence possess all properties derivable from that survey.

Our ``generic dynamical variable'' adopts the specific functional form $\mathcal{D}(\alpha, \beta) = \Vobs^\alpha / r^\beta$, allowing it to be fully parameterized by an angle $\theta=\arctan(\beta / \alpha)$. This choice is motivated by the fact that it encompasses all time derivatives of uniform circular motion---velocity, acceleration, jerk and higher---while offering a seamless interpolation between them. While one might consider parameterizing the dynamical space using a more intricate function, dimensional analysis alone imposes significant constraints. For instance, a linear combination of terms like $\mathcal{D}$ would require the addition of dimensionful coefficients, increasing complexity and difficulty of interpretation while limiting generalizability. A parallel argument applies to e.g. trigonometric functions or exponentials: their arguments must be dimensionless, necessitating a dimensionful constant to remove the dimensions of $\Vobs$ and $r$. Thus, while other dynamical variables could be possible, their further consideration is beyond the scope of this work.

There exists a range of $\tan \theta > 1 / 2$ over which $\mathcal{D}$ can be predicted equally well from several features ($\gbar,\, \Sigma_{\rm tot},\, D,\,i,\,L_{3.6},\,M_{\rm HI},\,T,\,R_{\rm eff},$ and $\e$). Our conservative primary conclusion is therefore that the \ac{RAR} represents the optimal \emph{1-dimensional} correlation between baryonic and dynamical properties. For all intents and purposes it is the optimal correlation of any dimension also, as the improvement observed in~\cref{fig:gencomb_ET_double} when using many features is statistically insignificant. While we cannot rule out the existence of a feature set that enhances predictivity of $\mathcal{D}$ beyond that of the \ac{RAR}, such evidence is not found in \ac{SPARC}.

Our analysis is not tied to a specific functional form for the \ac{RAR}: to benchmark the \ac{ML} regressor results we have employed the \ac{RARIF} and \ac{SIFEFE} simply as example $\gbar-\gobs$ relations that are known to fit the data reasonably well. This is fortunate, as~\citet{Desmond_2023_ESRRAR} demonstrated using symbolic regression that there exist more accurate and simpler functions than these. Utilizing the optimal functions from~\citeauthor{Desmond_2023_ESRRAR} would reduce the loss of the $\gbar-\gobs$ relation, but would not alter the conclusion that incorporating additional features does not improve accuracy. However, given the greater success of symbolic regression than numerical \ac{ML} methods it would be interesting to try to construct multi-variable symbolic fits for $\gobs$ using all available features. This would likely be less prone to overfitting when using an information-theoretic model selection criterion such as the minimum description length principle~\citep{ESR}, and to confusion when irrelevant features are considered. Feature importance may be assessed by the frequency of occurrence of variables in highly-ranked functions, or by the difference in accuracy or description length between similar functions including or not including a variable. While currently beyond the complexity limit of Exhaustive Symbolic Regression~\citep{ESR}, this could be achieved by genetic algorithms such as \textsc{Operon}~\citep{Operon}.


\subsubsection{Additional data}\label{sec:additional_data}

We have selected the \ac{SPARC} sample as representative of the late-type galaxy population, given its extensive range in morphology (S0 to Irr), luminosity and surface brightness, and unrestrictive selection criteria~\citep{SPARC}. To assess the impact of sample variance, we employed test-train splitting (a form of jackknifing) and Monte Carlo mock data generation, showing that our findings would be applicable to all \ac{SPARC}-like data realizations. However, to establish the \ac{RAR} as the fundamental dynamical correlation for late-type galaxies in general, further analysis on alternative data sets, such as LITTLE THINGS~\citep{iorio2017_littlethings} or PROBES~\citep{Stone2022}, is necessary. Although these larger data sets may yield increased statistical uncertainties due to lower quality requirements, the same conclusions should be reached if the error model is reliable.

Ideally, the results presented here would be generalized further to the dynamics of galaxies of all types. The \ac{RAR} has been investigated in diverse contexts, such as ultra-diffuse galaxies (e.g.~\citealt{Freundlich}), local dwarf spheroidals (e.g.~\citealt{McGaugh_Wolf, McGaugh_Milgrom}), low-acceleration regions including the outer Milky Way~\citep{Oman}, early-type galaxies~\citep{RAR,Shelest_2020,ellipticals_2, ellipticals_1, ellipticals_3} and galaxy groups and clusters~\citep{clusters_1, clusters_2, clusters_3, groups}. Although these data sets contain sufficient information for our analysis, the properties employed as features may differ, and additional systematic uncertainties may arise. However, given that the \ac{SPARC} galaxies are not qualitatively different to others---and are related to them evolutionarily---one would expect conclusions concerning the fundamentality of the \ac{RAR} to generalize. Exploring the \ac{RAR} using qualitatively different observables, such as stacked weak lensing~\citep{Brouwer}, could provide valuable consistency checks and extend our results to new regimes.


\subsubsection{Testing $\Lambda$CDM}

It would be interesting to perform our analysis on mock data generated by \ac{LCDM} (through a semi-analytic model and/or hydrodynamical simulation), to see if, as in the data, $\gobs$ is uniquely predictable from $\gbar$, $\{\alpha=2, r=1\}$ locates the $\Vobs^\alpha/r^\beta$ most strongly correlated with baryonic properties and the $\gbar-\gobs$ relation accounts for all dynamical correlations. It seems these would be unlikely coincidences, as neither $\gbar$ nor $\gobs$ have any particular significance in that theory.


\section{Conclusion}\label{sec:conc}
\acresetall

We explore the origin of galaxies' surprising simplicity by studying the \ac{RAR}, which generalizes and subsumes the statistical regularities of late-type galaxies' radial (in-disk) dynamics. In particular, we question whether the \ac{RAR} is \emph{the fundamental} relation of late-type radial dynamics, by which we mean that
\begin{enumerate}
    \item its residuals do not correlate significantly with any other quantity,
    \item it is the strongest dynamical-to-baryonic correlation,
    \item it is sufficient to explain all other correlations involving radial dynamics (in conjunction with the non-dynamical correlations of the data set we use), and
    \item no other relation possesses these properties.
\end{enumerate}
To do so we develop machine learning regressors for predicting dynamic variables from baryonic ones, establishing feature importance directly and by determining which variables contribute usefully to the prediction. We supplement this by a partial correlation analysis of \ac{RAR} residuals and validate our results on mock data that matches the non-dynamical properties of the \ac{SPARC} data while allowing control of its kinematic correlations.

We find the \ac{RAR} to satisfy all of these criteria for unique fundamentality: features besides $\gbar$ only degrade the prediction of $\gobs$ (which in fact cannot significantly exceed simple fitting functions in accuracy), predicting a general dynamical variable $\Vobs^\alpha/r^\beta$ picks out $\{\alpha=2, r=1\}$, i.e. $\gobs$, as the most predictable from baryonic features, and mock data including the \ac{RAR} but no other dynamical correlation has all $\{\alpha, \beta\}$ correlations approximately consistent with the real data. Subsidiary correlations such as between luminosity and velocity or surface brightness and jerk are explicable as projections of the \ac{RAR}, but not vice versa. Indeed, the second strongest $1$-dimensional dynamical correlation in the data possesses \emph{none} of the properties (i) - (iii) above. The \ac{RAR} therefore appears to be the fundamental $1$-dimensional correlation of late-type radial galaxy dynamics; consequently, to explain such dynamics it is necessary to explain the \ac{RAR}, and sufficient to explain the \ac{RAR} plus the lack of any other significant, partially independent correlation. This poses an extremely stringent test for theories of galaxy formation, which those based on modified dynamics readily pass while those based on concordance cosmology may not.


\section*{Data availability}

The \ac{SPARC} data is available at \url{http://astroweb.cwru.edu/SPARC/}. The code underlying this article is available at \url{https://github.com/Richard-Sti/RARinterpret} and other data will be made available on reasonable request to the authors.


\section*{Acknowledgements}

We thank Julien Devriendt, Harley Katz, Federico Lelli, Stacy McGaugh, Mordehai Milgrom, Adrianne Slyz and Tariq Yasin for useful comments and discussions. We also thank Jonathan Patterson for smoothly running the Glamdring Cluster hosted by the University of Oxford, where the data processing was performed. RS acknowledges financial support from STFC Grant No. ST/X508664/1. HD is supported by a Royal Society University Research Fellowship (grant no. 211046). This project has received funding from the European Research Council (ERC) under the European Union's Horizon 2020 research and innovation programme (grant agreement No 693024). For the purpose of open access, we have applied a Creative Commons Attribution (CC BY) licence to any Author Accepted Manuscript version arising.

\bibliographystyle{mnras}
\bibliography{ref}

\bsp
\label{lastpage}
\end{document}